# On guided circumferential waves in soft electroactive tubes under radially inhomogeneous biasing fields

Bin Wu[1,5], Yipin Su[1], Weiqiu Chen[1,2,3,4], Chuanzeng Zhang[5,*]

[1] Department of Engineering Mechanics, Zhejiang University, Hangzhou 310027, P.R. China

[2] State Key Laboratory of Fluid Power and Mechatronic Systems, Zhejiang University, Hangzhou 310027, P.R. China

[3] Key Laboratory of Soft Machines and Smart Devices of Zhejiang Province, Zhejiang University, Hangzhou 310027, P.R. China

[4] Soft Matter Research Center, Zhejiang University, Hangzhou 310027, P.R. China

[5] Department of Civil Engineering, University of Siegen, Siegen D-57068, Germany

**Abstract:**

Soft electroactive (EA) tube actuators and many other cylindrical devices have been proposed recently in literature, which show great advantages over those made from conventional hard solid materials. However, their practical applications may be limited because these soft EA devices are prone to various failure modes. In this paper, we present an analysis of the guided circumferential elastic waves in soft EA tube actuators, which has potential applications in the in-situ nondestructive evaluation or online structural health monitoring (SHM) to detect structural defects or fatigue cracks in soft EA tube actuators and in the self-sensing of soft EA tube actuators based on the concept of guided circumferential elastic waves. Both circumferential SH and Lamb-type waves in an incompressible soft EA cylindrical tube under inhomogeneous biasing fields are considered. The biasing fields, induced by the application of an electric voltage difference to the electrodes on the inner and outer cylindrical surfaces of the EA tube in addition to an axial pre-stretch, are inhomogeneous in the radial direction. Dorfmann and Ogden's theory of nonlinear electroelasticity and the associated linear theory for small incremental motion constitute the basis of our analysis. By means of the state-space formalism for the incremental wave motion along with the approximate laminate technique, dispersion relations are derived in a particularly efficient way. For a neo-Hookean ideal dielectric model, the proposed approach is first validated numerically. Numerical examples are then given to show that the guided circumferential wave propagation characteristics are significantly affected by the inhomogeneous biasing fields and the geometrical parameters. Some particular phenomena such as the frequency veering and the nonlinear dependence of the phase velocity on the radial electric voltage are discussed. Our numerical findings demonstrate that it is feasible to use guided circumferential elastic waves for the ultrasonic non-destructive online SHM to detect interior structural defects or fatigue cracks and for the self-sensing of the actual state of the soft EA tube actuator.

[*] Corresponding author. Tel.: +49 271 7402173; fax: +49 271 7404073.

E-mail address: c.zhang@uni-siegen.de (Ch. Zhang).

# 1. Introduction

Owing to the advantages such as low cost, light weight, rapid response and large deformation under electric stimuli, soft electroactive (EA) materials as one kind of smart materials have received considerable attention recently. Application of external electric fields can modify rapidly and reversibly the electromechanical properties of EA materials, prefiguring various potential applications including transducers, actuators, sensors, micro-pumps, biomedical devices, as well as flexible electronics such as displays, artificial muscles, and sensitive skins (Carpi et al., 2011; Anderson et al., 2012; Henann et al., 2013; Zhao and Wang, 2014). The high nonlinearity and notable electromechanical coupling characteristics of soft EA materials have prompted a lot of academic interest (Dorfmann and Ogden, 2005, 2006, 2014; McMeeking and Landis, 2005; Suo et al., 2008) in developing a general theoretical framework of nonlinear electroelasticity, which is more feasible and appropriate for describing the electromechanical behavior of soft EA materials such as dielectric elastomers (DEs) undergoing large deformation.

Although a wide range of novel applications of soft EA materials have been put forward, the high rates of failures due to the high-strain and high-voltage actuation limit their practical applications as commercial devices that require a long-term reliability and safety. Frequently, failures of the DE actuators originate from the electric failures due to the electromechanical instability induced pull-in effects (Zhao and Wang, 2014) or localized dielectric breakdown caused by interior defects present in DEs, which include inherent material defects caused during material processing such as gel particles, non-uniform cross-linking and foreign particulates, as well as defects induced during fabrication or actuation process such as puncture, voids, inclusions, mechanical stress concentration, electric field concentration and so on (Yuan et al., 2009; Stoyanov et al., 2013). In addition, the repeated stretching and contraction of the DE actuators represent a typical cyclic or fatigue loading and are prone to initiate fatigue cracks or damages at local imperfections or inhomogeneities in both the DE film and electrodes attached to its surfaces, and the fatigue damage evolution or crack growth may lead to the final failure of the entire EA devices under the in-service conditions (Lochmatter and Kovacs, 2008; Rajamani et al., 2008; Stoyanov et al., 2013). Specifically, Lochmatter and Kovacs (2008) determined experimentally the durability of an active hinge segment based on soft EA materials under cyclic actuation and pointed out that rather few actuation cycles were endured before a dielectric breakdown occurred. Rajamani et al. (2008) developed a DE wound roll actuator with high strain (12%) and stiffness (147 N/m) and tested cyclically several actuators at 3kV up to 3480 cycles until failure. By utilizing dielectric oil coated single-walled carbon nonotube electrodes, Yuan et al. (2009) reported a type of actuators which can be operated under a continuous large-strain actuation (150% in area) for longer than 1500 minutes without an ultimate failure. Long lifetime and fault-tolerant freestanding actuators based on a silicone dielectric elastomer and self-clearing carbon nanotube compliant electrodes were presented by Stoyanov et al. (2013) and their cyclic actuation tests indicated that the actuators still maintain a high level of performance for more than 85000 cycles at moderate electric fields. As pointed out by Lochmatter and Kovacs (2008) and Stoyanov et al. (2013), long-term actuations are possible but with a lower electromechanical performance than theoretically predicted, while a high electromechanical performance can be obtained at the expense of the lifetime. Therefore, the lifetime of soft EA materials significantly depends on both the mechanical and the electrical fatigues (Stoyanov et al., 2013).

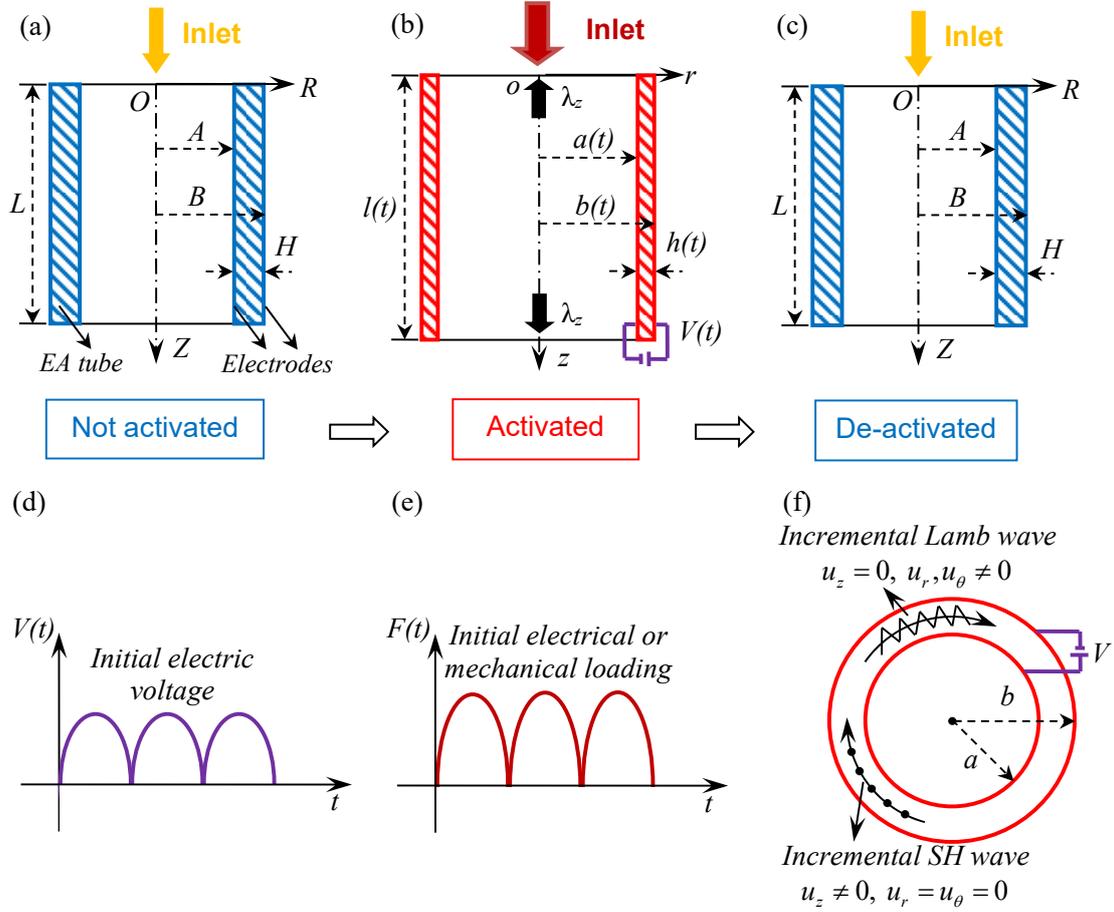

**Fig. 1.** ((a)-(c)) Schematic diagram of a soft EA tube with flexible electrodes: (a) undeformed (not activated) configuration; (b) deformed (or activated) configuration induced by an axial pre-stretch and a radial electric voltage; (c) de-activated configuration after removing the loading. (d) A particular periodic variation of the initial radial electric voltage versus time. (e) Cyclic initial mechanical or electrical loading induced by the cyclic initial radial electric voltage. (f) Two types of incremental guided circumferential waves superimposed on the deformed configuration corresponding to Fig. 1(b).

Since Pelrine et al. (1998) proposed for the first time the DE tube actuator, the particular configuration of the cylindrical soft EA tube actuators has attracted a great deal of attention (Carpi and Rossi, 2004; Rajamani et al., 2008; Son and Goulbourne, 2010; Zhu et al., 2010; Chen and Dai, 2012; Shmuel and deBotton, 2013; Zhou et al., 2014; Shmuel, 2015) from different points of view. Fig. 1(a) displays schematically an initially undeformed DE tube actuator which is bounded by flexible electrodes attached to its inner and outer surfaces. In this undeformed configuration, the EA tube is not activated, i.e., the tube is free of mechanical and electrical loads. Then an electric voltage difference is applied to the electrodes of the EA tube as shown in Fig. 1(b), which is also simultaneously subjected to an axial mechanical force with an axial pre-stretch. At this moment, the deformed EA tube is in an activated state and expands or squeezes axisymmetrically. By adjusting the electric voltage, different activated and deformed states may be reached. Once the EA tube is de-activated, i.e., the mechanical and electrical loads disappear, the actual state of the EA tube actuator recovers its original undeformed state as depicted in Fig. 1(c).

This process can be repeated and utilized to control the inlet and outlet of gaseous or liquid substances. A particular periodic variation of the applied initial radial electric voltage versus time is shown in Fig. 1(d), which leads to the repeated cyclic initial electrical and mechanical loading in the DE tube actuator (Fig. 1(e)). As a consequence, the DE tube actuator is more likely to sustain fatigue damages, defects, or cracks under this repeated cyclic loading, which may result in the mechanical or electrical fatigue failures of the DE tube actuator (Lochmatter and Kovacs, 2008; Rajamani et al., 2008; Stoyanov et al., 2013). For this reason, the real-time online structural health monitoring (SHM) to detect and characterize the interior structural defects or fatigue cracks in EA actuators is a crucial issue. For this purpose, the non-destructive ultrasonic techniques based on guided elastic waves, which have been widely used for monitoring the onset and progress of structural defects and fatigue cracks in linear elastic materials (Valle et al., 2001; Giurgiutiu, 2008), provide a possible and effective tool due to their particular characteristics of the propagation over long distance, multi-mode availability, sensitivity to different flaw types, and ability to follow the curvatures and reach the hidden or buried parts.

Guided circumferential waves, which can be effectively used to detect and characterize structural defects or fatigue cracks in cylindrical structures, have been an active research topic because of their pivotal significance not only in NDE (Valle et al., 2001; Luo et al., 2005) but also in the application realm of EA devices (White, 1970). For isotropic linear elastic materials, the investigation of Lamb-type or SH-type circumferential waves (hereafter abbreviated as Lamb waves or SH waves) propagating along the cylindrical surface have been conducted by Liu and Qu (1998), Gridin et al. (2003), and Zhao and Rose (2004), just to name a few. The guided circumferential waves propagating in anisotropic cylindrical curved plates were studied numerically by the Fourier series expansion technique (Towfighi et al., 2002). Chen (1973) investigated the SH wave propagation in a piezoelectric cylinder of hexagonal crystal symmetry. Yu et al. (2008, 2009) determined the circumferential wave propagation characteristics in electro-elastic and magneto-electro-elastic functionally graded cylindrical curved plates employing the Legendre orthogonal polynomial series expansion approach. Experimentally, measurements of circumferential guided waves have also been carried out by many researchers for the purpose of detecting defects (Valle et al., 2001; Luo et al., 2005) and extracting material properties (Nauleau et al., 2014; Lin et al., 2015; Chekroun et al., 2016). No biasing fields (such as initial stress, pre-stretch or pre-deformation, and biasing electric field) in the investigated structures were considered in the aforementioned works.

However, the performance of a multi-functional system made of EA materials is usually affected by various biasing fields since their presence may significantly alter its electromechanical properties (Yang and Hu, 2004; Dorfmann and Ogden, 2014). Therefore, a deep understanding of the physical phenomenon of guided ultrasonic waves propagating in soft EA materials under biasing fields is of paramount importance for in-situ nondestructive evaluation (NDE) or online SHM of soft EA actuators. Since the pioneering work on the theory of small dynamic fields superimposed on finite static biasing fields by Baumhauer and Tiersten (1973), the propagation of the small-amplitude elastic waves in EA materials under biasing fields has been a subject of intensive research interest (see a valuable review by Yang and Hu (2004) for a detailed survey). In particular, Chai and Wu (1996) used Lothe-Barnett's integral formalism to investigate the surface wave propagation characteristics in a pre-stressed piezoelectric material and proposed a possible application of the biasing fields to delay-controllable delay lines.

Lematre et al. (2006) employed the recursive stiffness matrix method proposed by Rokhlin and Wang (2002) to study the influence of the pre-stress gradient on the propagation of the Lamb and shear-horizontal (SH) waves in piezoelectric plates and surface acoustic waves in layered piezoelectric structures. However, most of the above-mentioned research works mainly focused on the effects of the biasing fields in the conventional hard piezoelectric materials. Recently, based on their nonlinear electroelasticity theory (Dorfmann and Ogden, 2005, 2006), Dorfmann and Ogden (2010) suggested a compact form of the linear incremental theory describing the small-amplitude motions superimposed on finite biasing fields, with a particular attention paid to the soft EA materials such as DEs.

As a practical application, the linear incremental theory by Dorfmann and Ogden (2010) lays a theoretical foundation for electrostatically tunable waveguides. Specifically, the linear incremental theory by Dorfmann and Ogden (2010) was applied to investigate the effects of the biasing fields on the propagation of the surface waves in a homogeneously deformed EA half-space (Dorfmann and Ogden, 2010), the generalized Rayleigh-Lamb waves in an ideal DE layer (Shmuel et al., 2012), the axisymmetric and non-axisymmetric waves in a pre-stretched incompressible EA cylinder and tube additionally subjected to an axial electric displacement (Chen and Dai, 2012; Su et al., 2016). All the above-mentioned works demonstrated that the wave propagation characteristics could be readily adjusted through the introduction of the biasing fields. In addition, they all assumed that the biasing fields are homogeneous such that exact solutions for various types of the elastic waves can be achieved. More recently, based on the neo-Hookean ideal dielectric model, Shmuel and deBotton (2013) and Shmuel (2015) considered the axisymmetric and torsional wave propagations in DE tubes under biasing fields induced by the combination of a radially applied electric field and an axial mechanical load, and showed that the DE tubes can be an appropriate candidate for active waveguides to tune elastic waves by means of proper biasing fields. Nonetheless, the application of an electric voltage difference to the compliant electrodes attached to the inner and outer surfaces of a cylindrical EA tube will result in radially inhomogeneous biasing fields, and hence it is intractable to obtain exact analytical solutions. Consequently, an effective numerical scheme with high accuracy should be established to overcome this difficulty.

Another important physical motivation to the present study is our self-sensing concept of soft EA actuators based on guided circumferential elastic waves. Since application of electric stimuli may result in significant and often complex changes of the wave propagation characteristics in the activated soft EA materials, it is possible to develop a novel self-sensing EA actuator based on guided waves in order to extract accurate actual information on the deformation (or other physical quantities) during the actuation process. Most of the existing self-sensing DE actuators operate by measuring the changes in the electrical characteristics of the DE actuators, such as the capacitance, electrode resistance, and dielectric resistance, due to the very large active deformation under an applied actuation voltage (Jung et al., 2008; Gisby et al., 2013; Hoffstadt et al., 2014). However, distinguished from these previous works, a self-sensing capability of an EA actuator can be achieved by using the measured variations of the voltage-dependent wave propagation characteristics such as the wave velocity and the wave mode. In this manner, the applied actuation voltage can be identified, controlled and self-adjusted if necessary, in order to precisely monitor and maintain the desired and stable actuation strain or deformation of the EA actuator.

In order to provide a theoretical guidance for applying the guided wave techniques to the online SHM to detect structural defects or fatigue cracks in activated soft EA tube actuators and to the self-sensing of soft EA tube actuators, we will examine the guided circumferential wave propagation characteristics in a soft EA tube actuator under inhomogeneous biasing fields in this paper. We first give in Section 2 a brief review of the general theory of nonlinear electroelasticity and the linear incremental theory for soft EA materials as formulated by Dorfmann and Ogden (2005, 2006, 2010, 2014). Without a specification for the energy function, we deal with the axisymmetric (or cylindrically symmetric) deformation of a soft EA tube with electrodes on its surfaces in Section 3. The EA tube is assumed to be subjected to an axial force or pre-stretch and an electric voltage difference applied to the two electrodes (the latter will be referred to as radial electric voltage afterwards for simplicity). Based on the state-space formalism for the incremental fields in cylindrical coordinates derived in Section 4, the approximate laminate technique is employed in Section 5 in order to efficiently obtain the dispersion relations for both the SH and Lamb waves propagating in the soft EA tube with underlying biasing fields that are radially inhomogeneous. For a neo-Hookean ideal dielectric model, the static axisymmetric large deformation of the tube and in particular the radially inhomogeneous biasing fields are illustrated graphically in Section 6.1. The proposed approach is verified in terms of its convergence and accuracy in Section 6.2. Numerical examples are finally presented to elucidate the SH and Lamb wave propagation characteristics in dependence on the biasing fields, including the axial pre-stretch and radial electric voltage, as well as the geometrical parameters in Sections 6.3 and 6.4, respectively. We draw some conclusions in Section 7 to summarize our main findings and discuss their further applications. The nomenclature and list of symbols used throughout this paper are cataloged in Appendix A. Some related detailed mathematical expressions or derivations are provided in the Appendices B-D.

**2. Basic formulations**

*2.1. Nonlinear electroelasticity theory*

The governing equations for small (incremental) fields superimposed on a finitely deformed configuration can be obtained from the general nonlinear electroelasticity theory, which will be briefly reviewed in this section. For a more detailed discussion about the basic ideas the interested readers are referred to the papers of Dorfmann and Ogden (2005, 2006, 2010) and the book of Dorfmann and Ogden (2014).

Consider a soft deformable EA continuum which, in the undeformed stress-free "reference configuration" at time $t_0$, occupies in the Euclidean space a region $\mathcal{B}_r$ with the boundary $\partial \mathcal{B}_r$ and the outward unit normal vector $\mathbf{N}$. The location of an arbitrary material point in this state is denoted by its position vector $\mathbf{X}$, which is a continuous labeling of that material point. At time $t$, the body occupies a region $\mathcal{B}_t$ with the boundary $\partial \mathcal{B}_t$ and the outward unit normal vector $\mathbf{n}_t$, if subjected to a motion $\mathbf{x} = \boldsymbol{\chi}(\mathbf{X},t)$, where $\boldsymbol{\chi}$ is a vector function with a sufficiently regular property. The current position of the material point associated with $\mathbf{X}$ is given by $\mathbf{x}$ and the region is referred to as "current configuration". The deformation gradient tensor is defined as $\mathbf{F} = \mathrm{Grad}\,\boldsymbol{\chi}$, where $\mathrm{Grad}$ is the gradient operator with respect to $\mathcal{B}_r$; and for its Cartesian components we have $\mathbf{F}_{i\alpha} = \partial x_i / \partial X_\alpha$. In this paper Greek indices are associated with the reference configuration $\mathcal{B}_r$ while Roman indices with $\mathcal{B}_t$, and the summation convention for repeated indices is adopted. The relations between the infinitesimal line element $\mathrm{d}\mathbf{X}$, surface element $\mathrm{d}A$ and volume element $dV$ in the undeformed configuration and those in the deformed configuration are specified by $d\mathbf{x} = \mathbf{F}d\mathbf{X}$, the well-known Nanson's formula $\mathbf{n}_t da_t = J\mathbf{F}^{-T}\mathbf{N}dA$, and $dv = JdV$, where $J = \det \mathbf{F}$ is a local measure of the volume change and the superscript T signifies transpose. Due to the incompressibility, we invariably have $J = 1$. The left and right Cauchy-Green strain tensors $\mathbf{b} = \mathbf{FF}^T$ and $\mathbf{C} = \mathbf{F}^T\mathbf{F}$ will be used as the deformation measures.

In the absence of free charges and electric currents, and with the quasi-electrostatic approximation, the Gauss's law and Faraday's law are given by

$$\text{div}\mathbf{D} = 0, \quad \text{curl}\mathbf{E} = \mathbf{0} \tag{1}$$

where $\mathbf{D}$ and $\mathbf{E}$ are the electric displacement and electric field vectors in $\mathcal{B}_t$; curl and div are the curl and divergence operators in $\mathcal{B}_t$, respectively. If the contribution of the electric body forces is included into the "total" Cauchy stress tensor denoted by $\boldsymbol{\tau}$, the equations of motion in the absence of the mechanical body forces may be written as

$$\text{div}\boldsymbol{\tau} = \rho \mathbf{x}_{,tt} \tag{2}$$

where $\rho$ is the material mass density, which does not change during the motion because of the material incompressibility, and the subscript $t$ following a comma denotes the material time derivative. The conservation of angular momentum leads to the symmetry of $\boldsymbol{\tau}$.

In this paper, we will consider an EA tube coated with electrodes on the inner and outer surfaces with free surface charges. In this case, as will be shown later, there is no electric field in the surrounding vacuum. Consequently, the electric boundary conditions to be satisfied on $\partial \mathcal{B}_t$ are

$$\mathbf{E} \times \mathbf{n}_t = \mathbf{0}, \quad \mathbf{D} \cdot \mathbf{n}_t = -\sigma_f \tag{3}$$

where $\sigma_f$ is the free surface charge density on $\partial \mathcal{B}_t$. In terms of the total Cauchy stress tensor, the mechanical boundary conditions may be written in Eulerian form as

$$\boldsymbol{\tau}\mathbf{n}_t = \mathbf{t}^a \tag{4}$$

where $\mathbf{t}^a$ is the applied mechanical traction vector per unit area of $\partial \mathcal{B}_t$. For the Lagrangian counterparts of Eqs. (1)-(4) we refer to the works of Dorfmann and Ogden (2005, 2006).

Following Dorfmann and Ogden (2005, 2006), the nonlinear constitutive relations for incompressible electroelastic materials in terms of the total energy density function $\Omega(\mathbf{F}, \mathcal{D})$ (per unit reference volume rather than per unit mass) are given by

$$\mathbf{T} = \frac{\partial \Omega}{\partial \mathbf{F}} - p\mathbf{F}^{-1}, \quad \boldsymbol{\mathcal{E}} = \frac{\partial \Omega}{\partial \mathcal{D}} \tag{5}$$

where $\mathbf{T} = \mathbf{F}^{-1}\boldsymbol{\tau}$, $\mathcal{D} = \mathbf{F}^{-1}\mathbf{D}$ and $\boldsymbol{\mathcal{E}} = \mathbf{F}^T\mathbf{E}$ are the "total" nominal stress tensor, the Lagrangian electric displacement and electric field vectors, respectively; $p$ is a Lagrange multiplier associated with the incompressibility constraint $J = 1$. For an incompressible isotropic electroelastic material, the total energy density function $\Omega$ can be reduced to a function depending on the following five invariants

$$I_1 = \text{tr}\mathbf{C}, \quad I_2 = [(\text{tr}\mathbf{C})^2 - \text{tr}(\mathbf{C}^2)]/2, \quad I_4 = \mathcal{D}\cdot\mathcal{D}, \quad I_5 = \mathcal{D}\cdot(\mathbf{C}\mathcal{D}), \quad I_6 = \mathcal{D}\cdot(\mathbf{C}^2\mathcal{D}) \quad (6)$$

Therefore, $\Omega = \Omega(I_1, I_2, I_4, I_5, I_6)$ and the explicit forms of the total stress tensor and the electric field vector are (Dorfmann and Ogden, 2010)

$$\boldsymbol{\tau} = 2\Omega_1 \mathbf{b} + 2\Omega_2(I_1\mathbf{b} - \mathbf{b}^2) - p\mathbf{I} + 2\Omega_5 \mathbf{D}\otimes\mathbf{D} + 2\Omega_6(\mathbf{D}\otimes\mathbf{bD} + \mathbf{bD}\otimes\mathbf{D}),$$
$$\mathbf{E} = 2(\Omega_4 \mathbf{b}^{-1}\mathbf{D} + \Omega_5 \mathbf{D} + \Omega_6 \mathbf{bD}) \quad (7)$$

where $\Omega_m = \partial\Omega/\partial I_m \; (m = 1, 2, 4, 5, 6)$.

*2.2. The linear incremental theory*

In this section, following Dorfmann and Ogden (2010), we present the governing equations for the time-dependent infinitesimal incremental changes in the motion and the electric displacement vector superimposed on an underlying static configuration $\mathcal{B}$ with the boundary $\partial\mathcal{B}$ and the outward unit normal vector $\mathbf{n}$ undergoing a finite deformation $\mathbf{x} = \chi(\mathbf{X})$ coupled with an electric displacement vector $\mathbf{D}(\mathbf{X})$. The current position of the material point associated with $\mathbf{X}$ is changed from $\mathbf{x}$ to $\mathbf{y}$ after the infinitesimal incremental motion. The concept of the three configurations, i.e., undeformed, initial, and current configurations, of the soft EA continuum is displayed in Fig. 2 for clarity. In principle, different coordinate systems may be employed to describe different configurations before and after deformations (Wu et al., 2016). Nonetheless, in this work the three coordinate systems are chosen to be coincident, i.e., their coordinate origins and base vectors are all $o$ and $\mathbf{i}_k$, and the results will not be affected. Here and henceforth a superposed dot indicates the incremental quantities rather than the temporal time derivative. However, the incremental displacement vector is represented by $\mathbf{u}$, not $\dot{\mathbf{u}}$, according to Dorfmann and Ogden (2010).

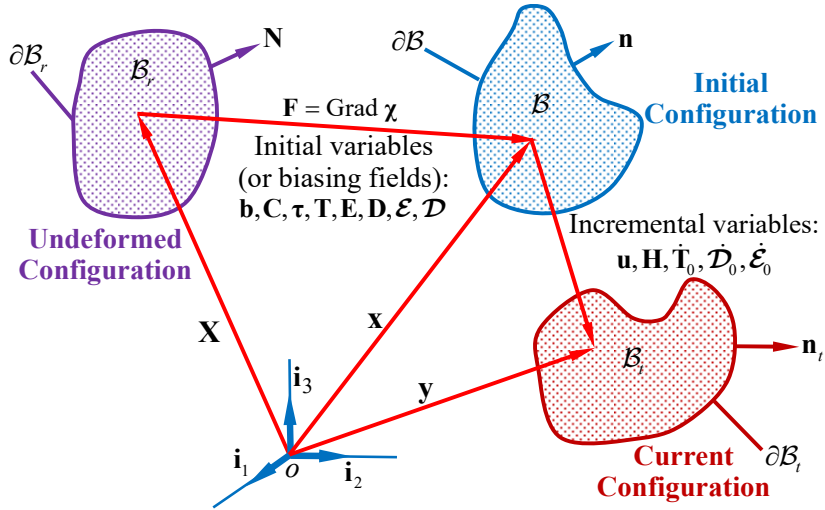

**Fig. 2.** The undeformed, initial, and current configurations of a soft EA continuum.

In Eulerian form, the incremental governing equations can be written as

$$\text{div}\dot{\mathcal{D}}_0 = 0, \quad \text{curl}\dot{\mathcal{E}}_0 = \mathbf{0}, \quad \text{div}\dot{\mathbf{T}}_0 = \rho \mathbf{u}_{,tt} \tag{8}$$

where $\dot{\mathbf{T}}_0$, $\dot{\mathcal{E}}_0$ and $\dot{\mathcal{D}}_0$ are the push-forward versions of the corresponding Lagrangian increments, which update the reference configuration from $\mathcal{B}_r$ to $\mathcal{B}$ (Dorfmann and Ogden, 2010). We identify the resulting "push-forward" variables with a subscript 0. The increments $\dot{\mathbf{T}}_0$ and $\dot{\mathcal{E}}_0$ satisfy the following linearized incremental constitutive laws for incompressible EA materials

$$\dot{\mathbf{T}}_0 = \mathcal{A}_0 \mathbf{H} + \mathcal{M}_0 \dot{\mathcal{D}}_0 + p\mathbf{H} - \dot{p}\mathbf{I}, \quad \dot{\mathcal{E}}_0 = \mathcal{M}_0^{\mathrm{T}} \mathbf{H} + \mathcal{R}_0 \dot{\mathcal{D}}_0 \tag{9}$$

where $\dot{p}$ is the incremental Lagrange multiplier and $\mathbf{H} = \text{grad}\,\mathbf{u}$ denotes the push-forward incremental displacement gradient tensor with grad being the gradient operator in $\mathcal{B}$. The instantaneous electroelastic moduli tensors $\mathcal{A}_0$, $\mathcal{M}_0$ and $\mathcal{R}_0$ are given in component notation by

$$\mathcal{A}_{0piqj} = F_{p\alpha} F_{q\beta} \mathcal{A}_{\alpha i \beta j} = \mathcal{A}_{0qjpi}, \quad \mathcal{M}_{0piq} = F_{p\alpha} F_{\beta q}^{-1} \mathcal{M}_{\alpha i \beta} = \mathcal{M}_{0ipq}, \quad \mathcal{R}_{0ij} = F_{\alpha i}^{-1} F_{\beta j}^{-1} \mathcal{R}_{\alpha\beta} = \mathcal{R}_{0ji} \tag{10}$$

where $\mathcal{A}$, $\mathcal{M}$ and $\mathcal{R}$ denote the referential electroelastic moduli tensors associated with $\Omega(\mathbf{F},\mathcal{D})$, with their components defined by $\mathcal{A}_{\alpha i \beta j} = \partial^2 \Omega / (\partial F_{i\alpha} \partial F_{j\beta})$, $\mathcal{M}_{\alpha i \beta} = \partial^2 \Omega / (\partial F_{i\alpha} \partial \mathcal{D}_\beta)$, and $\mathcal{R}_{\alpha\beta} = \partial^2 \Omega / (\partial \mathcal{D}_\alpha \partial \mathcal{D}_\beta)$. By using the incremental form of the symmetry condition $\mathbf{FT} = (\mathbf{FT})^{\mathrm{T}}$, the connections between the components of the tensors $\mathcal{A}_0$ and $\boldsymbol{\tau}$ for an incompressible material may be written as (Dorfmann and Ogden, 2014)

$$\mathcal{A}_{0jisk} - \mathcal{A}_{0ijsk} = (\tau_{js} + p\delta_{js})\delta_{ik} - (\tau_{is} + p\delta_{is})\delta_{jk} \tag{11}$$

In addition, the incremental incompressibility condition is given by

$$\text{div}\,\mathbf{u} = 0 \tag{12}$$

For an incompressible material, the Eulerian incremental forms of the electric and mechanical boundary conditions, which are to be satisfied on $\partial \mathcal{B}$, can be written as

$$\dot{\mathcal{E}}_0 \times \mathbf{n} = \mathbf{0}, \quad \dot{\mathcal{D}}_0 \cdot \mathbf{n} = -\dot{\sigma}_{\mathrm{F}0}, \quad \dot{\mathbf{T}}_0^{\mathrm{T}} \mathbf{n} = \dot{\mathbf{t}}_0^A \tag{13}$$

where the increments of electrical variables in the surrounding vacuum have been disregarded, $\dot{\sigma}_{\mathrm{F}0}$ and $\dot{\mathbf{t}}_0^A$ are the incremental surface charge and mechanical traction vector per unit area of $\partial \mathcal{B}$.

## 3. Axisymmetric deformation of a soft EA tube

The problem of the axisymmetric deformation of an EA tube immersed in a radial electric displacement field has been discussed by Singh and Pipkin (1966), Eringen and Maugin (1990), and Dorfmann and Ogden (2006). However, they all considered an EA tube without electrodes on its inner and outer surfaces. Although the axisymmetric deformation of an EA tube with a radial electric voltage applied to the compliant electrodes attached to its surfaces was recently investigated by Zhu et al. (2010), Shmuel and deBotton (2013), Shmuel (2015), and Zhou et al. (2014), they all focused on a particular material model such as the neo-Hookean or Gent model. In this section, based on the nonlinear electroelasticity theory described in Section 2.1, we analyze the axisymmetric deformation of an EA tube with electrodes on its inner and outer surfaces subjected to an axial force and a radial electric voltage, without the need to specify the energy density function.

As shown in Figs. 1(a) and 1(b), when the soft EA tube is in the undeformed configuration, its inner and outer radii are $A$ and $B$, and its length is $L$. Then an electric voltage difference $V$ is applied to the electrodes of the tube, which is also simultaneously subjected to an axial mechanical force with an axial pre-stretch $\lambda_z$. The inner and outer radii and the length of the deformed tube become $a$, $b$ and $l$, respectively. If the tube is taken to be incompressible, the axisymmetric deformation is described by (Dorfmann and Ogden 2006)

$$R^2 - A^2 = \lambda_z(r^2 - a^2), \quad \theta = \Theta, \quad z = \lambda_z Z \tag{14}$$

where $(R,\Theta,Z)$ and $(r,\theta,z)$ are the cylindrical coordinate systems in the undeformed and deformed configurations, respectively; $\lambda_z$ is the axial stretch, which is independent of $r$. Therefore, the deformation gradient tensor may be represented by a diagonal matrix $\mathbf{F} = \text{diag}[\lambda_r^{-1}\lambda_z^{-1}, \lambda_r, \lambda_z]$, where the radial and circumferential stretches are $\lambda_r^{-1}\lambda_z^{-1}$ and $\lambda_r = r/R$, respectively. Note that under the incompressibility constraint the deformation can be described by two independent stretches $\lambda_r$ and $\lambda_z$. For convenience, we introduce the following notational convention

$$H = B - A, \quad h = b - a, \quad \lambda_a = a/A, \quad \lambda_b = b/B, \quad \eta = A/B, \quad \bar{\eta} = a/b \tag{15}$$

Substituting Eqs. (15)$_{3,4}$ into Eq. (14), we obtain

$$\lambda_a^2 \lambda_z - 1 = \frac{R^2}{A^2}\left(\lambda_r^2 \lambda_z - 1\right) = \frac{B^2}{A^2}\left(\lambda_b^2 \lambda_z - 1\right) = \frac{1}{\eta^2}\left(\lambda_b^2 \lambda_z - 1\right) \tag{16}$$

The only non-zero component of the Eulerian electric displacement vector $\mathbf{D}$ is the radial component $D_r$ due to the axisymmetric deformation and the applied radial electric field. The Lagrangian electric displacement vector, $\mathcal{D} = \mathbf{F}^{-1}\mathbf{D}$, also has only one non-zero component $\mathcal{D}_r = \lambda_r \lambda_z D_r$. Consequently, in terms of the stretches and the radial electric displacement component, the five independent invariants in Eq. (6) can be written now as

$$I_1 = \lambda_r^{-2}\lambda_z^{-2} + \lambda_r^2 + \lambda_z^2, \quad I_2 = \lambda_r^2\lambda_z^2 + \lambda_r^{-2} + \lambda_z^{-2}, \quad I_4 = \lambda_r^2\lambda_z^2 D_r^2, \quad I_5 = \lambda_r^{-2}\lambda_z^{-2} I_4, \quad I_6 = \lambda_r^{-4}\lambda_z^{-4} I_4 \tag{17}$$

Based on the initial constitutive relations (7), the non-zero components of the total stress tensor $\boldsymbol{\tau}$ and the electric field vector $\mathbf{E}$ are obtained as

$$\begin{aligned}
\tau_{rr} &= 2\lambda_r^{-2}\lambda_z^{-2}\left[\Omega_1 + \Omega_2(\lambda_r^2 + \lambda_z^2)\right] + 2\left(\Omega_5 + 2\Omega_6 \lambda_r^{-2}\lambda_z^{-2}\right)D_r^2 - p, \\
\tau_{\theta\theta} &= 2\lambda_r^2\left[\Omega_1 + \Omega_2(\lambda_r^{-2}\lambda_z^{-2} + \lambda_z^2)\right] - p, \quad \tau_{zz} = 2\lambda_z^2\left[\Omega_1 + \Omega_2(\lambda_r^{-2}\lambda_z^{-2} + \lambda_r^2)\right] - p, \\
E_r &= 2(\Omega_4 \lambda_r^2 \lambda_z^2 + \Omega_5 + \Omega_6 \lambda_r^{-2}\lambda_z^{-2})D_r
\end{aligned} \tag{18}$$

From the expressions of the five invariants in Eq. (17), it is clear that only three independent quantities $\lambda_r$, $\lambda_z$ and $I_4$ remain and hence it is convenient to define a reduced energy function $\Omega^*$ as

$$\Omega^*(\lambda_r, \lambda_z, I_4) = \Omega(I_1, I_2, I_4, I_5, I_6), \tag{19}$$

Equations (17)-(19) lead to

$$\lambda_r \Omega^*_{\lambda_r} = \tau_{\theta\theta} - \tau_{rr}, \quad \lambda_z \Omega^*_{\lambda_z} = \tau_{zz} - \tau_{rr}, \quad E_r = 2\lambda_r^2 \lambda_z^2 \Omega^*_4 D_r \tag{20}$$

where $\Omega^*_{\lambda_r} = \partial \Omega^*/\partial \lambda_r$, $\Omega^*_{\lambda_z} = \partial \Omega^*/\partial \lambda_z$ and $\Omega^*_4 = \partial \Omega^*/\partial I_4$.

For the axisymmetric deformation, all physical quantities depend only on $r$. The Faraday's law (1)$_2$ is then satisfied automatically and the Gauss's law (1)$_1$ reduces to

$$\frac{1}{r}\frac{\partial(rD_r)}{\partial r}=0 \tag{21}$$

which means that $rD_r$ is a constant with $rD_r = aD_r(a) = bD_r(b)$, where $D_r(a)$ and $D_r(b)$ are the radial components of the electric displacement vector at the inner and outer surfaces. The EA tube is coated with compliant electrodes on both the inner and outer surfaces, with equal and opposite free surface charges $Q(a)$ and $Q(b)$, respectively, such that $Q(a)+Q(b)=0$. Then, there is no electric field outside the tube and the boundary condition (3)$_2$ gives $D_r(a) = \sigma_{fa}$ and $D_r(b) = -\sigma_{fb}$, where $\sigma_{fa}$ and $\sigma_{fb}$ are the free surface charge densities per unit deformed area $\partial \mathcal{B}$ on the inner and outer surfaces, respectively, defined by

$$\sigma_{fa} = \frac{Q(a)}{2\pi a \lambda_z L}, \quad \sigma_{fb} = \frac{Q(b)}{2\pi b \lambda_z L} \tag{22}$$

Thus, the solution of Eq. (21) may be written as

$$D_r = \frac{Q(a)}{2\pi r \lambda_z L} = -\frac{Q(b)}{2\pi r \lambda_z L} \tag{23}$$

Since the electric field is curl-free, it can be expressed in terms of the gradient of an electrostatic potential $\phi$, i.e., $\mathbf{E} = -\text{grad}\,\phi$. Then, the only non-zero electric field component is $E_r = -d\phi/dr$. Hence, with the help of Eqs. (20)$_3$ and (23), we have

$$\frac{d\phi}{dr} = -\lambda_r^2 \lambda_z \Omega_4^* \frac{Q(a)}{\pi r L} = \lambda_r^2 \lambda_z \Omega_4^* \frac{Q(b)}{\pi r L} \tag{24}$$

Denoting the electric voltage difference between the inner and outer surfaces by $V = \phi(a) - \phi(b)$, integrating Eq. (24) from the inner surface to the outer surface, we obtain

$$V = \lambda_z \frac{Q(a)}{\pi L} \int_a^b \lambda_r^2 \Omega_4^* \frac{dr}{r} = -\lambda_z \frac{Q(b)}{\pi L} \int_a^b \lambda_r^2 \Omega_4^* \frac{dr}{r} \tag{25}$$

which provides a general relation between the electric voltage difference and the surface free charge that depends on the initial deformation.

Since the deformation is axisymmetric, and in view of Eqs. (18) and (20)$_1$, the equilibrium equations $\text{div}\,\boldsymbol{\tau} = \mathbf{0}$ reduce to one equation only, i.e.,

$$\frac{d\tau_{rr}}{dr} = \frac{\lambda_r \Omega_{\lambda_r}^*}{r} \tag{26}$$

Since both the inner and outer surfaces of the EA tube are traction-free, i.e., $\tau_{rr}(a) = \tau_{rr}(b) = 0$, the integration of Eq. (26) from $a$ to $b$ gives

$$\int_a^b \lambda_r \Omega_{\lambda_r}^* \frac{dr}{r} = 0 \tag{27}$$

Note that $b$ may be expressed in terms of $a$ and $\lambda_z$ by Eq. (14)$_1$, and hence Eq. (27) establishes a general relation between the electrical variable ($V$ or $Q$), which is included in $\Omega^*$, and the inner radius $a$ when $\lambda_z$ is known. The radial normal stress can be found by integrating Eq. (26) from $a$ to $r$ as

$$\tau_{rr}(r) = \int_a^r \lambda_r \Omega^*_{\lambda_r} \frac{\mathrm{d}r}{r} \tag{28}$$

The integration of the axial normal stress $\tau_{zz}$ over the cross-section of the deformed EA tube gives the resultant axial force

$$N = 2\pi \int_a^b \tau_{zz}(r) r \mathrm{d}r \tag{29}$$

Making use of Eqs. (20)$_{1,2}$, the equilibrium equation (26) and the traction-free boundary conditions, the axial force can be rewritten as

$$N = \pi \int_a^b \left( 2\lambda_z \Omega^*_{\lambda_z} - \lambda_r \Omega^*_{\lambda_r} \right) r \mathrm{d}r \tag{30}$$

Therefore, once the reduced energy function $\Omega^*$ is determined, the integrations (25), (27)-(28) and (30) can be performed analytically or numerically. Then the circumferential normal stress $\tau_{\theta\theta}$ and the axial normal stress $\tau_{zz}$ can be obtained by Eqs. (20)$_1$ and (20)$_2$, respectively, after the radial normal stress $\tau_{rr}$ is determined from Eq. (28). The Lagrange multiplier $p$ can be determined by one of the three equations (18)$_{1\text{-}3}$. In Section 6.1, we will specialize the results obtained in this section to the neo-Hookean ideal dielectric model and present the explicit expressions.

## 4. State space formalism for the incremental fields

In this section, we will first recast the incremental governing equations obtained in Section 2.2 into their equivalent forms in the cylindrical coordinates $(r,\theta,z)$ in order to describe the time-dependent incremental motion and the accompanying incremental electric field in an EA tube. Then the state-space formalism for the incremental fields in the cylindrical coordinates will be derived.

In order to make the incremental Faraday's law (8)$_2$ satisfied identically, an incremental electric potential $\dot{\phi}$ can be introduced such that $\dot{\mathcal{E}}_0 = -\mathrm{grad}\dot{\phi}$, with the components in the cylindrical coordinates $(r,\theta,z)$ being

$$\dot{\mathcal{E}}_{0r} = -\frac{\partial \dot{\phi}}{\partial r}, \quad \dot{\mathcal{E}}_{0\theta} = -\frac{1}{r}\frac{\partial \dot{\phi}}{\partial \theta}, \quad \dot{\mathcal{E}}_{0z} = -\frac{\partial \dot{\phi}}{\partial z} \tag{31}$$

If the incremental displacements superimposed on the underlying deformed configuration as described in the previous section are denoted by $\mathbf{u} = u_r \mathbf{e}_r + u_\theta \mathbf{e}_\theta + u_z \mathbf{e}_z$, where $\mathbf{e}_r, \mathbf{e}_\theta, \mathbf{e}_z$ are the unit basis vectors in the cylindrical coordinates $(r,\theta,z)$ in the deformed configuration, the incremental displacement gradient tensor becomes

$$\mathbf{H} = \begin{bmatrix} \dfrac{\partial u_r}{\partial r} & \dfrac{1}{r}\left(\dfrac{\partial u_r}{\partial \theta} - u_\theta\right) & \dfrac{\partial u_r}{\partial z} \\ \dfrac{\partial u_\theta}{\partial r} & \dfrac{1}{r}\left(\dfrac{\partial u_\theta}{\partial \theta} + u_r\right) & \dfrac{\partial u_\theta}{\partial z} \\ \dfrac{\partial u_z}{\partial r} & \dfrac{1}{r}\dfrac{\partial u_z}{\partial \theta} & \dfrac{\partial u_z}{\partial z} \end{bmatrix} \tag{32}$$

and the incremental incompressibility constraint is given by

$$\frac{\partial u_r}{\partial r}+\frac{1}{r}\left(\frac{\partial u_\theta}{\partial \theta}+u_r\right)+\frac{\partial u_z}{\partial z}=0 \tag{33}$$

In the cylindrical coordinates, the corresponding incremental equations of motion $(8)_3$ and incremental Gauss's law $(8)_1$ can be written as

$$\begin{aligned}
\frac{\partial \dot{T}_{0rr}}{\partial r}+\frac{1}{r}\frac{\partial \dot{T}_{0\theta r}}{\partial \theta}+\frac{\dot{T}_{0rr}-\dot{T}_{0\theta\theta}}{r}+\frac{\partial \dot{T}_{0zr}}{\partial z}&=\rho\frac{\partial^2 u_r}{\partial t^2}\\
\frac{\partial \dot{T}_{0r\theta}}{\partial r}+\frac{1}{r}\frac{\partial \dot{T}_{0\theta\theta}}{\partial \theta}+\frac{\dot{T}_{0\theta r}+\dot{T}_{0r\theta}}{r}+\frac{\partial \dot{T}_{0z\theta}}{\partial z}&=\rho\frac{\partial^2 u_\theta}{\partial t^2}\\
\frac{\partial \dot{T}_{0rz}}{\partial r}+\frac{1}{r}\frac{\partial \dot{T}_{0\theta z}}{\partial \theta}+\frac{\partial \dot{T}_{0zz}}{\partial z}+\frac{\dot{T}_{0rz}}{r}&=\rho\frac{\partial^2 u_z}{\partial t^2}
\end{aligned} \tag{34}$$

and

$$\frac{\partial \dot{\mathcal{D}}_{0r}}{\partial r}+\frac{1}{r}\left(\frac{\partial \dot{\mathcal{D}}_{0\theta}}{\partial \theta}+\dot{\mathcal{D}}_{0r}\right)+\frac{\partial \dot{\mathcal{D}}_{0z}}{\partial z}=0 \tag{35}$$

For the axisymmetric deformation (14) of the EA tube subjected to an axial pre-stretch and a radial electric displacement field, the instantaneous electroelastic moduli tensors $\mathcal{A}_0$, $\mathcal{M}_0$ and $\mathcal{R}_0$ can be derived according to Dorfmann and Ogden (2010). The non-zero components of these tensors are given in Appendix A for the completeness of the presentation and, in addition, we have

$$\begin{aligned}
&\mathcal{A}_{0iijk}=0, \quad \mathcal{R}_{0jk}=0, \quad \text{for } j\neq k, \quad i\in\{1,2,3\}, \quad \text{no sum over } i,\\
&\mathcal{M}_{0ii2}=\mathcal{M}_{0ii3}=\mathcal{M}_{02ii}=\mathcal{M}_{03ii}=0, \quad i\in\{1,2,3\}, \quad \text{no sum over } i,\\
&\mathcal{M}_{0ijk}=0, \quad \text{for } i\neq j\neq k
\end{aligned} \tag{36}$$

It can be shown from Appendix A that the instantaneous electroelastic moduli depend on the applied mechanical pre-stretches $\lambda_r$ and $\lambda_z$, the biasing electric field $D_r$, and the specific form of the energy function $\Omega$. As a result, by adjusting the biasing fields, we can change the instantaneous electromechanical properties of the EA tube, which in turn results in paramount effects on the dynamic behavior of the incremental motions, such as wave propagation and vibration. Vice versa, by analyzing the measured changes in the wave propagation or vibration characteristics, the actual internal damage state of the EA tube induced by the repeated fatigue loading can be assessed, and the actual electromechanical properties of the EA tube can be self-sensed.

Consequently, using Eq. (36), the incremental constitutive equations (9) are reduced to

$$\begin{aligned}
\dot{T}_{0rr}&=(\mathcal{A}_{01111}+p)H_{11}+\mathcal{A}_{01122}H_{22}+\mathcal{A}_{01133}H_{33}+\mathcal{M}_{0111}\dot{\mathcal{D}}_{0r}-\dot{p},\\
\dot{T}_{0\theta\theta}&=\mathcal{A}_{01122}H_{11}+(\mathcal{A}_{02222}+p)H_{22}+\mathcal{A}_{02233}H_{33}+\mathcal{M}_{0221}\dot{\mathcal{D}}_{0r}-\dot{p},\\
\dot{T}_{0zz}&=\mathcal{A}_{01133}H_{11}+\mathcal{A}_{02233}H_{22}+(\mathcal{A}_{03333}+p)H_{33}+\mathcal{M}_{0331}\dot{\mathcal{D}}_{0r}-\dot{p},\\
\dot{T}_{0r\theta}&=\mathcal{A}_{01212}H_{21}+(\mathcal{A}_{01221}+p)H_{12}+\mathcal{M}_{0122}\dot{\mathcal{D}}_{0\theta},\\
\dot{T}_{0\theta r}&=\mathcal{A}_{02121}H_{12}+(\mathcal{A}_{01221}+p)H_{21}++\mathcal{M}_{0122}\dot{\mathcal{D}}_{0\theta},\\
\dot{T}_{0rz}&=\mathcal{A}_{01313}H_{31}+(\mathcal{A}_{01331}+p)H_{13}+\mathcal{M}_{0133}\dot{\mathcal{D}}_{0z},\\
\dot{T}_{0zr}&=\mathcal{A}_{03131}H_{13}+(\mathcal{A}_{01331}+p)H_{31}+\mathcal{M}_{0133}\dot{\mathcal{D}}_{0z},\\
\dot{T}_{0\theta z}&=\mathcal{A}_{02323}H_{32}+(\mathcal{A}_{02332}+p)H_{23},\\
\dot{T}_{0z\theta}&=\mathcal{A}_{03232}H_{23}+(\mathcal{A}_{02332}+p)H_{32},
\end{aligned} \tag{37}$$

and

$$\begin{aligned}
\dot{\mathcal{E}}_{0r} &= \mathcal{M}_{0111}H_{11} + \mathcal{M}_{0221}H_{22} + \mathcal{M}_{0331}H_{33} + \mathcal{R}_{011}\dot{\mathcal{D}}_{0r}, \\
\dot{\mathcal{E}}_{0\theta} &= \mathcal{M}_{0122}\left(H_{21} + H_{12}\right) + \mathcal{R}_{022}\dot{\mathcal{D}}_{0\theta}, \\
\dot{\mathcal{E}}_{0z} &= \mathcal{M}_{0133}\left(H_{31} + H_{13}\right) + \mathcal{R}_{033}\dot{\mathcal{D}}_{0z}
\end{aligned} \quad (38)$$

Solving for $\dot{\mathcal{D}}_0$ in Eq. (38) in terms of $\dot{\mathcal{E}}_0$, then substituting the resulting expressions into Eq. (37) and taking account of the relations (31) and (32), we can transform the incremental constitutive equations (37) and (38) into those in terms of the incremental electric potential $\dot{\phi}$ and displacement vector **u** as

$$\begin{aligned}
\dot{T}_{0rr} &= c_{11}\frac{\partial u_r}{\partial r} + c_{12}\frac{1}{r}\left(\frac{\partial u_\theta}{\partial \theta} + u_r\right) + c_{13}\frac{\partial u_z}{\partial z} + e_{11}\frac{\partial \dot{\phi}}{\partial r} - \dot{p}, \\
\dot{T}_{0\theta\theta} &= c_{12}\frac{\partial u_r}{\partial r} + c_{22}\frac{1}{r}\left(\frac{\partial u_\theta}{\partial \theta} + u_r\right) + c_{23}\frac{\partial u_z}{\partial z} + e_{12}\frac{\partial \dot{\phi}}{\partial r} - \dot{p}, \\
\dot{T}_{0zz} &= c_{13}\frac{\partial u_r}{\partial r} + c_{23}\frac{1}{r}\left(\frac{\partial u_\theta}{\partial \theta} + u_r\right) + c_{33}\frac{\partial u_z}{\partial z} + e_{13}\frac{\partial \dot{\phi}}{\partial r} - \dot{p}, \\
\dot{T}_{0\theta z} &= c_{44}\frac{1}{r}\frac{\partial u_z}{\partial \theta} + c_{47}\frac{\partial u_\theta}{\partial z}, \quad \dot{T}_{0z\theta} = c_{47}\frac{1}{r}\frac{\partial u_z}{\partial \theta} + c_{77}\frac{\partial u_\theta}{\partial z}, \\
\dot{T}_{0rz} &= c_{55}\frac{\partial u_z}{\partial r} + c_{58}\frac{\partial u_r}{\partial z} + e_{35}\frac{\partial \dot{\phi}}{\partial z}, \quad \dot{T}_{0zr} = c_{58}\frac{\partial u_z}{\partial r} + c_{88}\frac{\partial u_r}{\partial z} + e_{35}\frac{\partial \dot{\phi}}{\partial z}, \\
\dot{T}_{0r\theta} &= c_{66}\frac{\partial u_\theta}{\partial r} + c_{69}\frac{1}{r}\left(\frac{\partial u_r}{\partial \theta} - u_\theta\right) + e_{26}\frac{1}{r}\frac{\partial \dot{\phi}}{\partial \theta}, \\
\dot{T}_{0\theta r} &= c_{69}\frac{\partial u_\theta}{\partial r} + c_{99}\frac{1}{r}\left(\frac{\partial u_r}{\partial \theta} - u_\theta\right) + e_{26}\frac{1}{r}\frac{\partial \dot{\phi}}{\partial \theta}
\end{aligned} \quad (39)$$

and

$$\begin{aligned}
\dot{\mathcal{D}}_{0r} &= e_{11}\frac{\partial u_r}{\partial r} + e_{12}\frac{1}{r}\left(\frac{\partial u_\theta}{\partial \theta} + u_r\right) + e_{13}\frac{\partial u_z}{\partial z} - \varepsilon_{11}\frac{\partial \dot{\phi}}{\partial r}, \\
\dot{\mathcal{D}}_{0\theta} &= e_{26}\left[\frac{1}{r}\left(\frac{\partial u_r}{\partial \theta} - u_\theta\right) + \frac{\partial u_\theta}{\partial r}\right] - \varepsilon_{22}\frac{1}{r}\frac{\partial \dot{\phi}}{\partial \theta}, \\
\dot{\mathcal{D}}_{0z} &= e_{35}\left(\frac{\partial u_z}{\partial r} + \frac{\partial u_r}{\partial z}\right) - \varepsilon_{33}\frac{\partial \dot{\phi}}{\partial z}
\end{aligned} \quad (40)$$

where

$$\begin{aligned}
&\varepsilon_{11} = 1/\mathcal{R}_{011}, \quad \varepsilon_{22} = 1/\mathcal{R}_{022}, \quad \varepsilon_{33} = 1/\mathcal{R}_{033}, \quad e_{11} = -\mathcal{M}_{0111}\varepsilon_{11}, \\
&e_{12} = -\mathcal{M}_{0221}\varepsilon_{11}, \quad e_{13} = -\mathcal{M}_{0331}\varepsilon_{11}, \quad e_{26} = -\mathcal{M}_{0122}\varepsilon_{22}, \quad e_{35} = -\mathcal{M}_{0133}\varepsilon_{33}, \\
&c_{11} = \mathcal{A}_{01111} + \mathcal{M}_{0111}e_{11} + p, \quad c_{12} = \mathcal{A}_{01122} + \mathcal{M}_{0111}e_{12}, \quad c_{13} = \mathcal{A}_{01133} + \mathcal{M}_{0111}e_{13}, \\
&c_{22} = \mathcal{A}_{02222} + \mathcal{M}_{0221}e_{12} + p, \quad c_{23} = \mathcal{A}_{02233} + \mathcal{M}_{0331}e_{12}, \quad c_{33} = \mathcal{A}_{03333} + \mathcal{M}_{0331}e_{13} + p, \quad c_{44} = \mathcal{A}_{02323}, \\
&c_{47} = \mathcal{A}_{02332} + p, \quad c_{77} = \mathcal{A}_{03232}, \quad c_{55} = \mathcal{A}_{01313} + \mathcal{M}_{0133}e_{35}, \quad c_{58} = \mathcal{A}_{01331} + \mathcal{M}_{0133}e_{35} + p, \\
&c_{88} = \mathcal{A}_{03131} + \mathcal{M}_{0133}e_{35}, \quad c_{66} = \mathcal{A}_{01212} + \mathcal{M}_{0122}e_{26}, \quad c_{69} = \mathcal{A}_{01221} + \mathcal{M}_{0122}e_{26} + p, \quad c_{99} = \mathcal{A}_{02121} + \mathcal{M}_{0122}e_{26}
\end{aligned} \quad (41)$$

It can be found from Eqs. (39)-(41) that the application of biasing fields makes the originally isotropic EA material behave like a conventional anisotropic piezoelectric material with respect to the underlying deformed configuration. This symmetry-breaking corresponds to the so-called "deformation-induced anisotropy" (Yang and Hu, 2004; Wu et al., 2016). This phenomenon has also been noticed by Su et al. (2016) for an isotropic EA hollow cylinder subjected to an axial pre-stretch and an axial electric displacement.

In the displacement-based method, the incremental stresses and the incremental electric displacements are usually eliminated from Eqs. (31)-(35) and (39)-(40) to obtain four coupled second-order partial differential equations for the incremental displacements and electric potential. However, the application of a radial electric voltage makes the biasing fields inhomogeneous. Specifically, the biasing fields as described in Section 3 generally are functions of $r$, leading to the $r$-dependence of the instantaneous electroelastic moduli as given in Appendix A. Therefore, the resulting incremental "displacement" equations in general are a system of coupled partial differential equations with variable coefficients, which are difficult to solve analytically or numerically. In Section 5, in order to overcome this difficulty and study the effects of the inhomogeneous biasing fields on the circumferential waves in the deformed EA tube, we will present an efficient method, which combines the state-space formalism with the approximate laminate technique.

In contrast to the conventional displacement-based method, the state-space method (SSM) as a special mixed-variables method usually uses three stress components, three displacement components, the electric potential and one electric displacement component as basic variables (e.g. the state variables), and transforms the governing equations into a set of first-order ordinary differential equations with respect to one particular coordinate variable, the radial coordinate here. The SSM has several particular advantages over the displacement-based method in solving many practical problems, and the interested readers are referred to Ding and Chen (2001) and Chen and Ding (2012) for more details and the references cited therein.

Following a standard way (Ding and Chen, 2001), the state equation can be readily derived from Eqs. (33)-(35) and (39)-(40). For simplicity, the detailed derivation procedure has been omitted here and we directly give the following final state equation

$$\frac{\partial \mathbf{Y}}{\partial r} = \mathbf{M} \mathbf{Y} \tag{42}$$

where the incremental state vector $\mathbf{Y}$ is defined as

$$\mathbf{Y} = \begin{bmatrix} u_r & u_\theta & u_z & \dot{\phi} & \dot{T}_{0rr} & \dot{T}_{0r\theta} & \dot{T}_{0rz} & \dot{\mathcal{D}}_{0r} \end{bmatrix}^{\mathrm{T}} \tag{43}$$

and $\mathbf{M}$ is the $8 \times 8$ system matrix, with its four partitioned $4 \times 4$ sub-matrices being given in Appendix B.

Note that the state equation (42) can be applied without any restrictions on the specific form of the energy density function. In addition to the state equation (42), which can be efficiently solved, when supplemented with appropriate boundary conditions as shown in the next section, a set of output equations is usually needed for the determination of the physical variables other than the state variables in Eq. (43). Nonetheless, for our purpose in this paper, these are not necessary and hence disregarded.

## 5. Dispersion relations of the circumferential waves in an EA tube

Since guided circumferential waves can be readily utilized to detect and characterize structural defects or fatigue damages in cylindrical structures (Liu and Qu, 1998; Valle et al., 2001; Zhao and Rose, 2004; Luo et al., 2005), in this section, we will consider a small-amplitude circumferential wave motion in an EA tube subjected to the biasing fields determined in Section 3. Based on the state-space formalism obtained in the previous section, the approximate laminate technique will be used here to derive the dispersion relations of the circumferential waves. Since we are mainly interested in the circumferential elastic waves propagating around the periphery of the EA tube without a variation in $z$, we have the relation $\partial/\partial z = 0$ and the state equation (42) can be simplified considerably. In fact, by setting $\partial/\partial z = 0$ in Eq. (42) and rearranging the state variables in Eq. (43), we have

$$\frac{\partial \mathbf{Y}_k}{\partial r} = \mathbf{M}_k \mathbf{Y}_k, \quad k \in \{1, 2\} \tag{44}$$

where $\mathbf{Y}_1 = \left[ u_z, \dot{T}_{0rz} \right]^{\mathrm{T}}$, $\mathbf{Y}_2 = \left[ u_r, u_\theta, \dot{\phi}, \dot{T}_{0rr}, \dot{T}_{0r\theta}, \dot{\mathcal{D}}_{0r} \right]^{\mathrm{T}}$ and

$$\mathbf{M}_1 = \begin{bmatrix} 0 & \dfrac{1}{c_{55}} \\ \rho\dfrac{\partial^2}{\partial t^2} - \dfrac{c_{44}}{r^2}\dfrac{\partial^2}{\partial \theta^2} & -\dfrac{1}{r} \end{bmatrix},$$

$$\mathbf{M}_2 = \begin{bmatrix} -\dfrac{1}{r} & -\dfrac{1}{r}\dfrac{\partial}{\partial \theta} & 0 & 0 & 0 & 0 \\ -\dfrac{c_{69}}{c_{66}}\dfrac{1}{r}\dfrac{\partial}{\partial \theta} & \dfrac{c_{69}}{c_{66}}\dfrac{1}{r} & -\dfrac{e_{26}}{c_{66}}\dfrac{1}{r}\dfrac{\partial}{\partial \theta} & 0 & \dfrac{1}{c_{66}} & 0 \\ \dfrac{q_1}{r} & \dfrac{q_1}{r}\dfrac{\partial}{\partial \theta} & 0 & 0 & 0 & -\dfrac{1}{\varepsilon_{11}} \\ \rho\dfrac{\partial^2}{\partial t^2} - \dfrac{q_7}{r^2}\dfrac{\partial^2}{\partial \theta^2} + \dfrac{q_3}{r^2} & \dfrac{q_3 + q_7}{r^2}\dfrac{\partial}{\partial \theta} & -\dfrac{q_8}{r^2}\dfrac{\partial^2}{\partial \theta^2} & 0 & -\dfrac{c_{69}}{c_{66}}\dfrac{1}{r}\dfrac{\partial}{\partial \theta} & -\dfrac{q_1}{r} \\ -\dfrac{q_3 + q_7}{r^2}\dfrac{\partial}{\partial \theta} & \rho\dfrac{\partial^2}{\partial t^2} - \dfrac{q_3}{r^2}\dfrac{\partial^2}{\partial \theta^2} + \dfrac{q_7}{r^2} & -\dfrac{q_8}{r^2}\dfrac{\partial}{\partial \theta} & -\dfrac{1}{r}\dfrac{\partial}{\partial \theta} & -\left(\dfrac{c_{69}}{c_{66}} + 1\right)\dfrac{1}{r} & \dfrac{q_1}{r}\dfrac{\partial}{\partial \theta} \\ -\dfrac{q_8}{r^2}\dfrac{\partial^2}{\partial \theta^2} & \dfrac{q_8}{r^2}\dfrac{\partial}{\partial \theta} & \dfrac{q_{11}}{r^2}\dfrac{\partial^2}{\partial \theta^2} & 0 & -\dfrac{e_{26}}{c_{66}}\dfrac{1}{r}\dfrac{\partial}{\partial \theta} & -\dfrac{1}{r} \end{bmatrix} \tag{45}$$

It is clear from Eqs. (44) and (45) that the eight state variables in Eq. (43) have been divided into two groups of independent physical variables, which means that there exist two types of incremental circumferential waves superimposed on the underlying deformed configuration as depicted in Fig. 1(f): the incremental SH waves described by $\mathbf{Y}_1$ and $\mathbf{M}_1$, whose mechanical displacement vector is out of the $r-\theta$-plane (anti-plane) of the deformed EA tube and parallel to the $z$-axis (i.e., $u_z \neq 0$, $u_r = u_\theta = 0$), and the incremental Lamb waves governed by $\mathbf{Y}_2$ and $\mathbf{M}_2$, whose mechanical displacement vector lies entirely in the $r-\theta$-plane and perpendicular to the $z$-axis (i.e., $u_z = 0$, $u_r, u_\theta \neq 0$).

For time-harmonic circumferential waves, we assume traveling wave solutions as

$$\begin{aligned} u_r &= bU_r(\xi)\exp[\mathrm{i}(\nu\theta - \omega t)], \quad u_\theta = bU_\theta(\xi)\exp[\mathrm{i}(\nu\theta - \omega t)], \\ u_z &= bU_z(\xi)\exp[\mathrm{i}(\nu\theta - \omega t)], \quad \dot{\phi} = b\sqrt{\mu/\varepsilon}\,\Phi(\xi)\exp[\mathrm{i}(\nu\theta - \omega t)], \\ \dot{T}_{0rr} &= \mu\Sigma_{0rr}(\xi)\exp[\mathrm{i}(\nu\theta - \omega t)], \quad \dot{T}_{0r\theta} = \mu\Sigma_{0r\theta}(\xi)\exp[\mathrm{i}(\nu\theta - \omega t)], \\ \dot{T}_{0rz} &= \mu\Sigma_{0rz}(\xi)\exp[\mathrm{i}(\nu\theta - \omega t)], \quad \dot{\mathcal{D}}_{0r} = \sqrt{\mu\varepsilon}\,\Delta_{0r}(\xi)\exp[\mathrm{i}(\nu\theta - \omega t)] \end{aligned} \tag{46}$$

where $\xi = r/b$ is the dimensionless radial coordinate and $\mu$ denotes the shear modulus of the EA material in the absence of an electric field; $\nu$ and $\omega$ are the angular wave number and circular frequency, respectively; $\varepsilon$ denotes the dielectric constant of the ideal dielectric material to be considered in Section 6, which is independent of the deformation. Usually, the angular wave number should be an integer to guarantee the periodicity in $\theta$ for steady-state wave solutions. However, for traveling circumferential waves, the time interval considered or the width of the wave pulses is chosen in such a way that the steady-state has not been reached. Therefore, the angular wave number is not necessarily an integer (Chen, 1973).

In order to ensure the circumferential waves to have a plane wave front, the linear phase velocity, denoted by $c_r$, at a distance $r$ from the center should be proportional to the radius $r$, i.e., $c_r = c_b r/b = c_b \xi$, where $c_b$ is the propagating phase velocity at the outer surface of the deformed EA tube which can be measured easily (Liu and Qu, 1998). Therefore, in contrast to the flat surface case, the linear wave number of the curved tube defined by $k_r = \omega/c_r$ is $r$ dependent. However, by defining the angular phase velocity as $\alpha = c_r/r = c_b/b$, the angular wave number, which is defined as $\nu = \omega/\alpha$, is independent of $r$ (Towfighi et al., 2002). Therefore, the linear phase velocity of the circumferential waves at a distance $r$ from the center is determined by

$$c_r = r\alpha = r\omega/\nu \tag{47}$$

Especially, the linear phase velocity at the outer surface of the deformed EA tube is given by

$$c_b = b\alpha = b\omega/\nu \tag{48}$$

In fact, Eq. (46) represents a solution which has the same phase factor along the radial lines and each radial line of the EA tube can be regarded as a wave front of the circumferential waves (Liu and Qu, 1998).

Substituting Eq. (46) into Eqs. (44) and (45) yields

$$\frac{d}{d\xi}\mathbf{V}_k(\xi) = \bar{\mathbf{M}}_k \mathbf{V}_k(\xi), \quad k \in \{1, 2\} \tag{49}$$

where $\mathbf{V}_1 = [U_z, \Sigma_{0rz}]^T$, $\mathbf{V}_2 = [U_r, U_\theta, \Phi, \Sigma_{0rr}, \Sigma_{0r\theta}, \Delta_{0r}]^T$ and the non-dimensional system matrices $\bar{\mathbf{M}}_1$ and $\bar{\mathbf{M}}_2$ are given by

$$\bar{\mathbf{M}}_1 = \begin{bmatrix} 0 & \dfrac{\mu}{c_{55}} \\ -\varpi^2 s^2 + \dfrac{c_{44}}{\mu}\dfrac{\nu^2}{\xi^2} & -\dfrac{1}{\xi} \end{bmatrix}, \tag{50}$$

and

$$\bar{\mathbf{M}}_2 = \begin{bmatrix} -\dfrac{1}{\xi} & -\dfrac{i\nu}{\xi} & 0 & 0 & 0 & 0 \\ -\beta_1\dfrac{i\nu}{\xi} & \dfrac{\beta_1}{\xi} & -\beta_2\dfrac{i\nu}{\xi} & 0 & \dfrac{\mu}{c_{66}} & 0 \\ \dfrac{\beta_3}{\xi} & \beta_3\dfrac{i\nu}{\xi} & 0 & 0 & 0 & -\dfrac{\varepsilon}{\varepsilon_{11}} \\ -\varpi^2 s^2 + \dfrac{\nu^2}{\xi^2}\dfrac{q_7}{\mu} + \dfrac{1}{\xi^2}\dfrac{q_3}{\mu} & \dfrac{i\nu}{\xi^2}\dfrac{q_3+q_7}{\mu} & \beta_4\dfrac{\nu^2}{\xi^2} & 0 & -\beta_1\dfrac{i\nu}{\xi} & -\dfrac{\beta_3}{\xi} \\ -\dfrac{i\nu}{\xi^2}\dfrac{q_3+q_7}{\mu} & -\varpi^2 s^2 + \dfrac{\nu^2}{\xi^2}\dfrac{q_3}{\mu} + \dfrac{1}{\xi^2}\dfrac{q_7}{\mu} & -\beta_4\dfrac{i\nu}{\xi^2} & -\dfrac{i\nu}{\xi} & -\dfrac{\beta_1+1}{\xi} & \beta_3\dfrac{i\nu}{\xi} \\ \dfrac{\nu^2}{\xi^2}\beta_4 & \dfrac{i\nu}{\xi^2}\beta_4 & -\dfrac{\nu^2}{\xi^2}\dfrac{q_{11}}{\varepsilon} & 0 & -\beta_2\dfrac{i\nu}{\xi} & -\dfrac{1}{\xi} \end{bmatrix} \quad (51)$$

in which we have introduced the following dimensionless quantities

$$s = \frac{b}{H} = \frac{\lambda_b}{1-\eta}, \quad \beta_1 = \frac{c_{69}}{c_{66}}, \quad \beta_2 = \frac{e_{26}}{c_{66}}\sqrt{\frac{\mu}{\varepsilon}}, \quad \beta_3 = q_{11}\sqrt{\frac{\varepsilon}{\mu}}, \quad \beta_4 = \frac{q_8}{\sqrt{\mu\varepsilon}} \quad (52)$$

and

$$\varpi = \frac{\omega H}{c_T} \quad (53)$$

where $c_T = \sqrt{\mu/\rho}$ is the shear wave velocity in the EA tube without initial biasing fields. Taking into account of Eqs. (47)-(48) and (52)-(53), we can express the linear phase velocity of the circumferential waves as

$$c_r = \xi s \frac{\varpi c_T}{\nu}, \quad c_b = s \frac{\varpi c_T}{\nu} \quad (54)$$

As described in Sections 3 and 4, the biasing fields are inhomogeneous in the radial direction, which makes the instantaneous electroelastic moduli tensors dependent on the radial coordinate $r$. Therefore, it is evident from Eqs. (50) and (51) that the system matrices $\bar{\mathbf{M}}_1$ and $\bar{\mathbf{M}}_2$ vary with $\xi$, which makes it troublesome to get the exact solutions to Eq. (49) directly. For this reason, the approximate laminate or multi-layer model (Thomson, 1950; Fan and Zhang, 1992; Chen and Ding, 2002; Chen et al., 2004) is employed in this paper to obtain the approximate analytical solutions. For this purpose, the deformed EA tube is equally divided into $n$ thin layers, each with a sufficiently small thickness $h/n$, such that the system matrices $\bar{\mathbf{M}}_1$ and $\bar{\mathbf{M}}_2$ within each layer may be assumed approximately as constant rather than variable. Consequently, the solutions in the $j$th layer can be obtained as (Thomson, 1950; Fan and Zhang, 1992; Ding and Chen, 2001; Chen and Ding, 2002; Chen et al., 2004; Chen and Ding, 2012)

$$\left.\begin{array}{l}\mathbf{V}_1(\xi) = \exp[(\xi-\xi_{j0})\bar{\mathbf{M}}_{1j}(\xi_{jm})]\mathbf{V}_1(\xi_{j0}) \\ \mathbf{V}_2(\xi) = \exp[(\xi-\xi_{j0})\bar{\mathbf{M}}_{2j}(\xi_{jm})]\mathbf{V}_2(\xi_{j0})\end{array}\right\}, \quad \left(\xi_{j0} \leq \xi \leq \xi_{j1}, \; j=1,2,3,\cdots n\right) \quad (55)$$

where $\bar{\mathbf{M}}_{1j}(\xi_{jm})$ and $\bar{\mathbf{M}}_{2j}(\xi_{jm})$ denote the approximated system matrices for the SH and Lamb waves, respectively, which are constant within the $j$th layer by taking $\xi = \xi_{jm}$; $\xi_{j0}$, $\xi_{j1}$ and $\xi_{jm}$ are the dimensionless radial coordinates at the inner, outer and middle surfaces of the $j$th layer, respectively, i.e.,

$$\xi_{j0} = \bar{\eta} + (j-1)\frac{1-\bar{\eta}}{n}, \quad \xi_{j1} = \bar{\eta} + j\frac{1-\bar{\eta}}{n}, \quad \xi_{jm} = \bar{\eta} + \frac{(2j-1)(1-\bar{\eta})}{2n} \quad (56)$$

in which $\xi_{10} = \bar{\eta}$ and $\xi_{n1} = 1$. Setting $\xi = \xi_{j1}$ in Eq. (55) gives rise to

$$\left.\begin{array}{l} \mathbf{V}_1(\xi_{j1}) = \exp[(1-\bar{\eta})\bar{\mathbf{M}}_{1j}/n]\mathbf{V}_1(\xi_{j0}) \\ \mathbf{V}_2(\xi_{j1}) = \exp[(1-\bar{\eta})\bar{\mathbf{M}}_{2j}/n]\mathbf{V}_2(\xi_{j0}) \end{array}\right\} \quad (57)$$

which represent the relations between the state vectors at the inner and outer surfaces of the $j$th layer. The continuity conditions at each fictitious interface of the layers require the eight state variables be continuous. Thus, we obtain from Eq. (57)

$$\mathbf{V}_k^1 = \mathbf{K}_k \mathbf{V}_k^0, \quad k \in \{1, 2\} \quad (58)$$

where $\mathbf{K}_k = \prod_{j=n}^{1} \exp[(1-\bar{\eta})\bar{\mathbf{M}}_{1k}/n]$ is the global transfer matrix of second-order ($k=1$) or sixth-order ($k=2$); and $\mathbf{V}_k^0$ and $\mathbf{V}_k^1$ are the state vectors at the inner and outer surfaces, respectively.

For the incremental motion of the EA tube, there are also no incremental electric fields outside the tube. Furthermore, if we assume that there is no incremental mechanical traction and that the applied electric voltage difference keeps unchanged during the incremental motion, the corresponding mechanical and electric boundary conditions (13) reduce to

$$\Sigma_{0rr}^0 = \Sigma_{0r\theta}^0 = \Sigma_{0rz}^0 = \Phi^0 = \Sigma_{0rr}^1 = \Sigma_{0r\theta}^1 = \Sigma_{0rz}^1 = \Phi^1 = 0 \quad (59)$$

Applying the incremental boundary conditions (59) in Eq. (58) results in two independent dispersion equations

$$K_{121} = 0, \quad \begin{vmatrix} K_{231} & K_{232} & K_{236} \\ K_{241} & K_{242} & K_{246} \\ K_{251} & K_{252} & K_{256} \end{vmatrix} = 0 \quad (60)$$

where $K_{kij}$ are the elements of the matrix $\mathbf{K}_k$. Equations (60)$_{1,2}$ determine the dispersion relations between the angular wave number $\nu$ and the circular frequency $\varpi$ for the incremental SH and Lamb waves, respectively. Once the dispersion equations (60) are solved for the $\varpi - \nu$ relations, the propagating phase velocity along the outer surface can be obtained from Eq. (54)$_2$.

## 6. Numerical results and discussions

### *6.1. The neo-Hookean ideal dielectric model*

For numerical illustration, the EA material is assumed to be characterized by the following neo-Hookean ideal dielectric model

$$\Omega = \mu(I_1 - 3)/2 + I_5/(2\varepsilon) \quad (61)$$

Utilizing Eq. (17)$_{1,4}$, the energy function (61) can be written in the reduced form as

$$\Omega^*(\lambda_r, \lambda_z, I_4) = \mu(\lambda_r^{-2}\lambda_z^{-2} + \lambda_r^2 + \lambda_z^2 - 3)/2 + \lambda_r^{-2}\lambda_z^{-2}I_4/(2\varepsilon) \quad (62)$$

For the neo-Hookean model (61) or (62), Zhu et al. (2010) have obtained the explicit expressions of the essential physical variables for the axisymmetric deformation, in particular the radially inhomogeneous biasing fields in the soft EA tube. We cite their results below, but use our notation. First, the equations governing the nonlinear response of the soft EA tube can be written as

$$\bar{\eta} = \lambda_a \eta \lambda_b^{-1}, \quad \bar{V} = -\frac{\eta \bar{Q}}{1-\eta} \ln \bar{\eta},$$

$$\bar{V} = -\sqrt{\lambda_z^{-1}\left(2\frac{\lambda_a^2}{1-\bar{\eta}^2}\ln\frac{\lambda_a}{\lambda_b} + \lambda_a^2 - \lambda_z^{-1}\right)} \frac{\eta}{1-\eta} \ln\bar{\eta}, \quad (63)$$

$$\bar{N} = \pi\left[\frac{(\lambda_z^2 - \lambda_z^{-1})\lambda_b^2(1-\bar{\eta}^2)}{(1-\eta)^2} + \frac{\lambda_z^{-1}\eta^2(\lambda_a^2 - \lambda_z^{-1})\ln\eta}{(1-\eta)^2} + \frac{2\lambda_z^{-1}\lambda_a^2}{(1-\bar{\eta}^2)(\eta^{-1}-1)^2}\ln\frac{\lambda_a}{\lambda_b}\ln\bar{\eta}\right]$$

where the dimensionless electric potential difference, surface charge, and resultant axial force are defined as

$$\bar{V} = \frac{V}{H}\sqrt{\frac{\varepsilon}{\mu}}, \quad \bar{Q} = \frac{Q(a)}{2\pi A \lambda_z L \sqrt{\mu\varepsilon}}, \quad \bar{N} = \frac{N}{\mu H^2} \quad (64)$$

We note that the expression for the resultant axial force in Zhu et al. (2010) contains typographical errors, which has been corrected in Eq. (63)$_4$. Furthermore, the radially inhomogeneous biasing fields are given by

$$\lambda_r = \frac{\xi}{\sqrt{\eta^2 \lambda_b^{-2} + \lambda_z(\xi^2 - \bar{\eta}^2)}}, \quad \bar{D}_r = \frac{\eta}{\lambda_b \xi}\bar{Q} = -\frac{1-\eta}{\lambda_b \xi \ln \bar{\eta}}\bar{V},$$

$$\bar{\tau}_{rr} = \lambda_z^{-1}\left[\ln\frac{\lambda_a}{\lambda_r} + \frac{\bar{\eta}^2 - \xi^2}{(1-\bar{\eta}^2)\xi^2}\ln\frac{\lambda_a}{\lambda_b}\right], \quad \bar{\tau}_{\theta\theta} = \lambda_z^{-1}\left[\ln\frac{\lambda_a}{\lambda_r} - \frac{\bar{\eta}^2 + \xi^2}{(1-\bar{\eta}^2)\xi^2}\ln\frac{\lambda_a}{\lambda_b} + \lambda_z \lambda_r^2 - 1\right], \quad (65)$$

$$\bar{\tau}_{zz} = \lambda_z^{-1}\left[\ln\frac{\lambda_a}{\lambda_r} - \frac{\bar{\eta}^2 + \xi^2}{(1-\bar{\eta}^2)\xi^2}\ln\frac{\lambda_a}{\lambda_b} + \lambda_z^3 - 1\right], \quad \bar{p} = \lambda_z^{-1}\left[1 - \ln\frac{\lambda_a}{\lambda_r} + \frac{\bar{\eta}^2 + \xi^2}{(1-\bar{\eta}^2)\xi^2}\ln\frac{\lambda_a}{\lambda_b}\right]$$

where the dimensionless quantities in Eq. (65) are defined as

$$\bar{D}_r = \frac{D_r}{\sqrt{\mu\varepsilon}}, \quad \bar{\tau}_{rr} = \frac{\tau_{rr}}{\mu}, \quad \bar{\tau}_{\theta\theta} = \frac{\tau_{\theta\theta}}{\mu}, \quad \bar{\tau}_{zz} = \frac{\tau_{zz}}{\mu}, \quad \bar{p} = \frac{p}{\mu} \quad (66)$$

As a matter of fact, substituting Eq. (62) into the general expressions in Section 3 and integrating the resulting equations lead to Eqs. (63)-(66), as expected. This serves as a means to check the correctness of these formulae.

In the absence of an electric voltage, i.e., $\bar{V} = 0$, and when the EA tube is only subjected to a mechanical pre-stretch $\lambda_z$, the deformation will be homogeneous, with the radial and circumferential stretches being $\lambda_z^{-1/2}$ along with $\bar{\tau}_{rr} = \bar{\tau}_{\theta\theta} = 0$. Correspondingly, by substituting Eq. (62) into Eq. (20)$_2$ and setting $D_r = 0$, we obtain the axial normal stress $\bar{\tau}_{zz} = \lambda_z^2 - \lambda_z^{-1}$, and hence the axial force $\bar{N} = \pi(1-\eta^2)(\lambda_z - \lambda_z^{-2})/(1-\eta)^2$.

For the simplified energy function (61) or (62), the non-zero components of the instantaneous electroelastic moduli tensors given in Appendix A can be evaluated as

$$\mathcal{A}_{01111} = \mathcal{A}_{01212} = \mathcal{A}_{01313} = \mu \lambda_r^{-2} \lambda_z^{-2} + \varepsilon^{-1} D_r^2, \quad \mathcal{A}_{02121} = \mathcal{A}_{02222} = \mathcal{A}_{02323} = \mu \lambda_r^2,$$

$$\mathcal{A}_{03131} = \mathcal{A}_{03232} = \mathcal{A}_{03333} = \mu \lambda_z^2, \quad \mathcal{M}_{0111} = 2\varepsilon^{-1} D_r, \quad \mathcal{M}_{0122} = \mathcal{M}_{0133} = \varepsilon^{-1} D_r, \quad \mathcal{R}_{011} = \mathcal{R}_{022} = \mathcal{R}_{033} = \varepsilon^{-1} \quad (67)$$

Thus the material parameters defined in Eq. (41) become

$$\varepsilon_{11} = \varepsilon_{22} = \varepsilon_{33} = \varepsilon, \quad e_{12} = e_{13} = 0, \quad e_{11} = -2D_r, \quad e_{26} = e_{35} = -D_r,$$
$$c_{11} = \mu\lambda_r^{-2}\lambda_z^{-2} - 3\varepsilon^{-1}D_r^2 + p, \quad c_{22} = \mu\lambda_r^2 + p, \quad c_{12} = c_{13} = c_{23} = 0, \quad c_{33} = \mu\lambda_z^2 + p,$$
$$c_{44} = \mu\lambda_r^2, \quad c_{55} = c_{66} = \mu\lambda_r^{-2}\lambda_z^{-2}, \quad c_{77} = \mu\lambda_z^2, \quad c_{88} = \mu\lambda_r^2 - \varepsilon^{-1}D_r^2, \quad (68)$$
$$c_{99} = \mu\lambda_z^2 - \varepsilon^{-1}D_r^2, \quad c_{47} = p, \quad c_{58} = c_{69} = p - \varepsilon^{-1}D_r^2$$

From Appendix B, we obtain

$$q_1 = 2\varepsilon^{-1}D_r, \quad q_3 = \varepsilon^{-1}D_r^2 + \mu(\lambda_r^2 + \lambda_r^{-2}\lambda_z^{-2}) + 2p, \quad q_7 = \mu\lambda_r^2 - \varepsilon^{-1}D_r^2 - \mu^{-1}(p - \varepsilon^{-1}D_r^2)^2\lambda_r^2\lambda_z^2,$$
$$q_8 = -D_r\left[1 + (\mu^{-1}\varepsilon^{-1}D_r^2 - \mu^{-1}p)\lambda_r^2\lambda_z^2\right], \quad q_{11} = \varepsilon + \mu^{-1}D_r^2\lambda_r^2\lambda_z^2 \quad (69)$$

As a consequence, the dimensionless quantities defined in Eq. (52) can be rewritten as

$$\beta_1 = (\bar{p} - \bar{D}_r^2)\lambda_r^2\lambda_z^2, \quad \beta_2 = -\bar{D}_r\lambda_r^2\lambda_z^2, \quad \beta_3 = 2\bar{D}_r, \quad \beta_4 = -\bar{D}_r\left[(\bar{D}_r^2 - \bar{p})\lambda_r^2\lambda_z^2 + 1\right] \quad (70)$$

and the dimensionless quantities appearing in Eq. (51) become

$$\frac{\mu}{c_{55}} = \frac{\mu}{c_{66}} = \lambda_r^2\lambda_z^2, \quad \frac{c_{44}}{\mu} = \lambda_r^2, \quad \frac{\varepsilon}{\varepsilon_{11}} = 1, \quad \frac{q_3}{\mu} = \lambda_r^2 + \lambda_r^{-2}\lambda_z^{-2} + \bar{D}_r^2 + 2\bar{p},$$
$$\frac{q_7}{\mu} = \lambda_r^2 - (\bar{p} - \bar{D}_r^2)^2\lambda_r^2\lambda_z^2 - \bar{D}_r^2, \quad \frac{q_{11}}{\varepsilon} = 1 + \bar{D}_r^2\lambda_r^2\lambda_z^2 \quad (71)$$

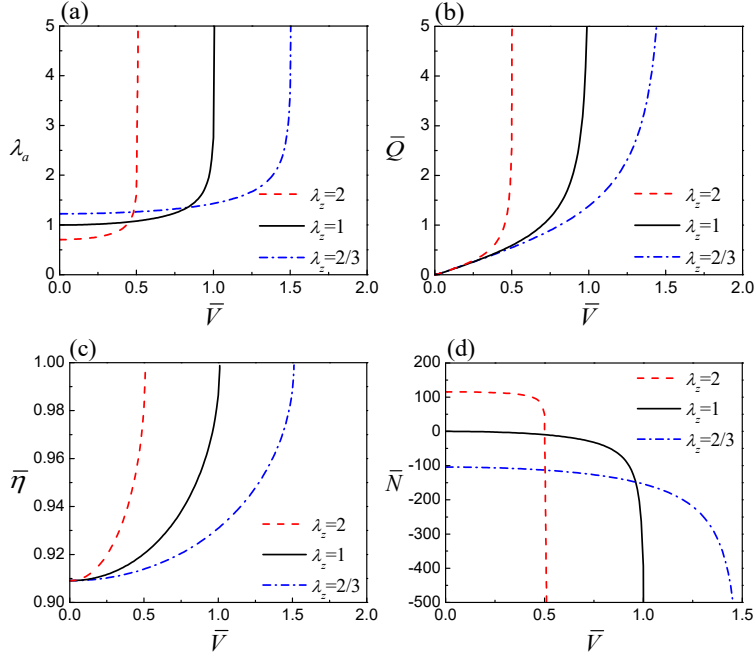

**Fig. 3.** Variations of $\lambda_a$, $\bar{Q}$, $\bar{\eta}$ and $\bar{N}$ with $\bar{V}$ for a thin EA tube at different pre-stretches $\lambda_z$ where $\bar{V} = 0.51$, 1.01 and 1.51 are the critical voltages beyond that the EA tube will collapse.

Based on Eq. (63), the dimensionless quantities $\lambda_a$, $\bar{Q}$, $\bar{\eta}$ and $\bar{N}$ versus the dimensionless electric voltage $\bar{V}$ are shown in Fig. 3 for different axial pre-stretches $\lambda_z$, where the ratio of the inner radius to the outer radius in the undeformed configuration has been set as $\eta = A/B = 1/1.1$, which corresponds to a quite thin tube. It can be seen from Fig. 3 that, beyond a critical voltage $\bar{V}_c$, which depends on the geometry of the EA tube and the axial pre-stretch, no solution of the axisymmetric deformation exists and the EA tube collapses (Shmuel and deBotton, 2013; Shmuel, 2015). The critical voltages of the thin tube are about 0.51, 1.01 and 1.51 for $\lambda_z = 2$, 1 and 2/3, respectively. Furthermore, it is noted from Figs. 3(a)-3(c) that, for a fixed pre-stretch, $\lambda_a$, $\bar{\eta}$ and $\bar{Q}$ increase with increasing $\bar{V}$, which means physically that the tube expands, its thickness decreases, and the surface charge accumulates on the electrodes as a result of the applied electric voltage. In addition, when the electric voltage approaches the critical voltage $\bar{V}_c$, $\lambda_a$ and $\bar{Q}$ have a tremendous rise and the dimensionless axial force decreases sharply in a nonlinear way. The results for a thick EA tube are qualitatively similar to those of a thin EA tube, and hence they are not shown here for the sake of brevity. The main difference is that for a thick EA tube, a slightly higher critical voltage is required to reach the steeply rising/decreasing segments. For example, the critical voltages for a thick EA tube with $\eta = A/B = 1/5$ are calculated to be about 0.56, 1.11 and 1.66 for $\lambda_z = 2$, 1 and 2/3, respectively.

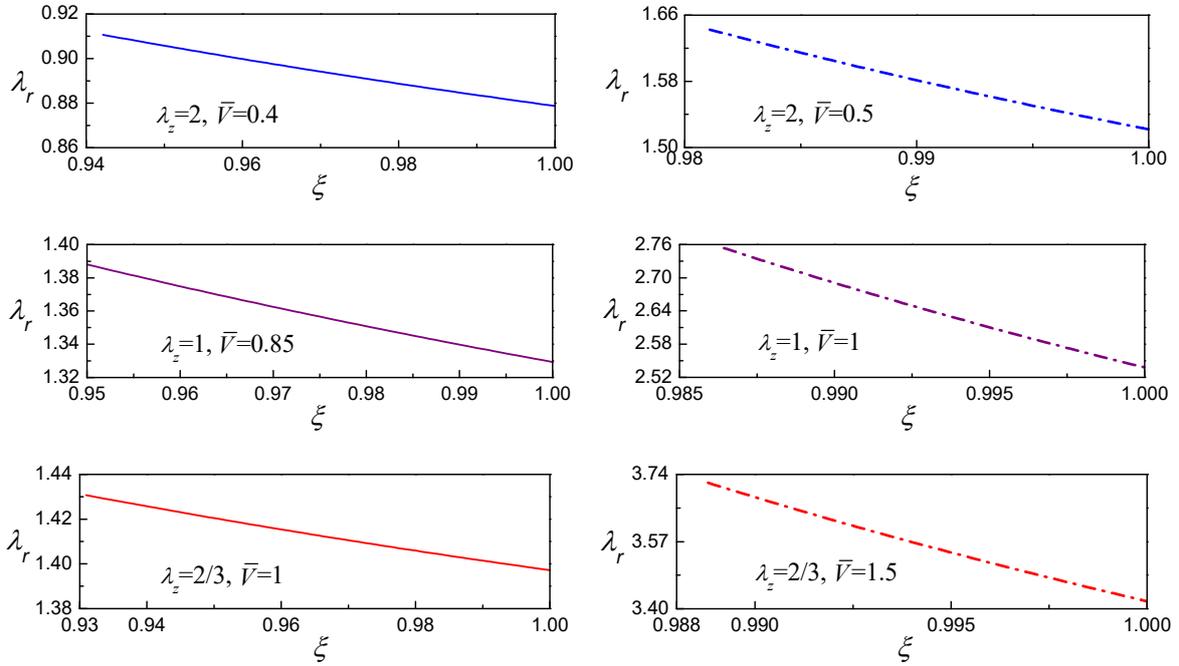

**Fig. 4.** Radial distribution of $\lambda_r$ in a thin EA tube for different combinations of $\lambda_z$ and $\bar{V}$.

According to Eq. (65)$_1$, we plot the radial distribution of the circumferential stretch $\lambda_r$ of the thin EA tube ($\eta = 1/1.1$) in Fig. 4 for different combinations of the pre-stretch $\lambda_z$ and the electric voltage $\bar{V}$. Note that the starting point $\bar{\eta}$ of the dimensionless radial coordinate $\xi$ in the deformed configuration is different for different pre-stretch and electric voltage. As mentioned previously, in the absence of an electric voltage, the deformation of the pre-stretched EA tube is homogeneous. However, when an electric voltage is applied, the biasing fields become radially inhomogeneous. It can be observed in Fig. 4 that, the circumferential stretch $\lambda_r$ is inhomogeneous and decreases approximately linearly with the increasing dimensionless radial coordinate $\xi$. Specifically, for a fixed pre-stretch, the inhomogeneity of the circumferential stretch is insignificant under a low electric voltage, whereas the degree of the inhomogeneity increases substantially when the electric voltage tends to the critical value. Besides, the circumferential stretch $\lambda_r$ at the inner surface is always larger than that at the outer surface of the EA tube, as shown in Fig. 4, which is obviously consistent with Eq. (16).

The radial distributions of $\bar{D}_r$, $\bar{\tau}_{rr}$, $\bar{\tau}_{\theta\theta}$, $\bar{\tau}_{zz}$ and $\bar{p}$ determined by Eqs. (65)$_{2-6}$ are also inhomogeneous. Since these inhomogeneities are well inferable from that of $\lambda_r$, they are not shown here for the sake of brevity. Also, for a thick EA tube, the results are similar to those of a thin EA tube, and hence they are omitted here as well.

*6.2. Validation of the analysis*

In this section, we will verify the analysis based on the state-space formalism along with the approximate laminate technique (referred to as SSM for brevity hereafter) in terms of the convergence and accuracy for studying the circumferential waves in an EA tube.

Tables 1 and 2 give the first two lowest dimensionless frequencies $\varpi$ of the SH and Lamb waves, respectively, in a thick EA tube ($\eta = 1/5$) with $\lambda_z = 1$ and $\bar{V} = 1.1$ for two dimensionless angular wave numbers $\nu = 5$ and 15 calculated using different numbers of the discretized layers *n*. It can be seen that the SSM has an excellent convergence rate. In fact, when the layer number increases, the approximate laminate model will gradually approach the original EA tube that is radially inhomogeneous. Therefore, accurate numerical results with an arbitrary precision may be obtained by the present SSM.

**Table 1** The first two lowest dimensionless frequencies of SH waves in a thick EA tube calculated by the SSM with different numbers of the discretized layers ($\lambda_z = 1$ and $\bar{V} = 1.1$)

| *n* | 20 | 40 | 60 | 80 | 100 | 120 | 140 | 160 |
|---|---|---|---|---|---|---|---|---|
| $\nu = 5$ | 5.13415 | 5.1329 | 5.13266 | 5.13258 | 5.13254 | 5.13252 | 5.13251 | 5.1325 |
| | 8.41829 | 8.41568 | 8.41519 | 8.41502 | 8.41494 | 8.4149 | 8.41488 | 8.41487 |
| $\nu = 15$ | 13.6231 | 13.618 | 13.617 | 13.6167 | 13.6165 | 13.6164 | 13.6164 | 13.6164 |
| | 17.7188 | 17.715 | 17.7144 | 17.7141 | 17.714 | 17.7139 | 17.7139 | 17.7139 |

**Table 2** The first two lowest dimensionless frequencies of Lamb waves in a thick EA tube calculated by the SSM with different numbers of the discretized layers ($\lambda_z = 1$ and $\bar{V} = 1.1$)

| *n* | 20 | 40 | 60 | 80 | 100 | 120 | 140 | 160 |
|---|---|---|---|---|---|---|---|---|
| $\nu = 5$ | 9.67882 | 9.67467 | 9.67391 | 9.67364 | 9.67352 | 9.67345 | 9.67343 | 9.67342 |
| | 17.3153 | 17.3135 | 17.3132 | 17.3131 | 17.313 | 17.313 | 17.313 | 17.3129 |
| $\nu = 15$ | 9.2096 | 9.22028 | 9.22251 | 9.22331 | 9.22369 | 9.22389 | 9.22491 | 9.22492 |
| | 28.6662 | 28.6512 | 28.6484 | 28.6475 | 28.647 | 28.6468 | 28.6467 | 28.6466 |

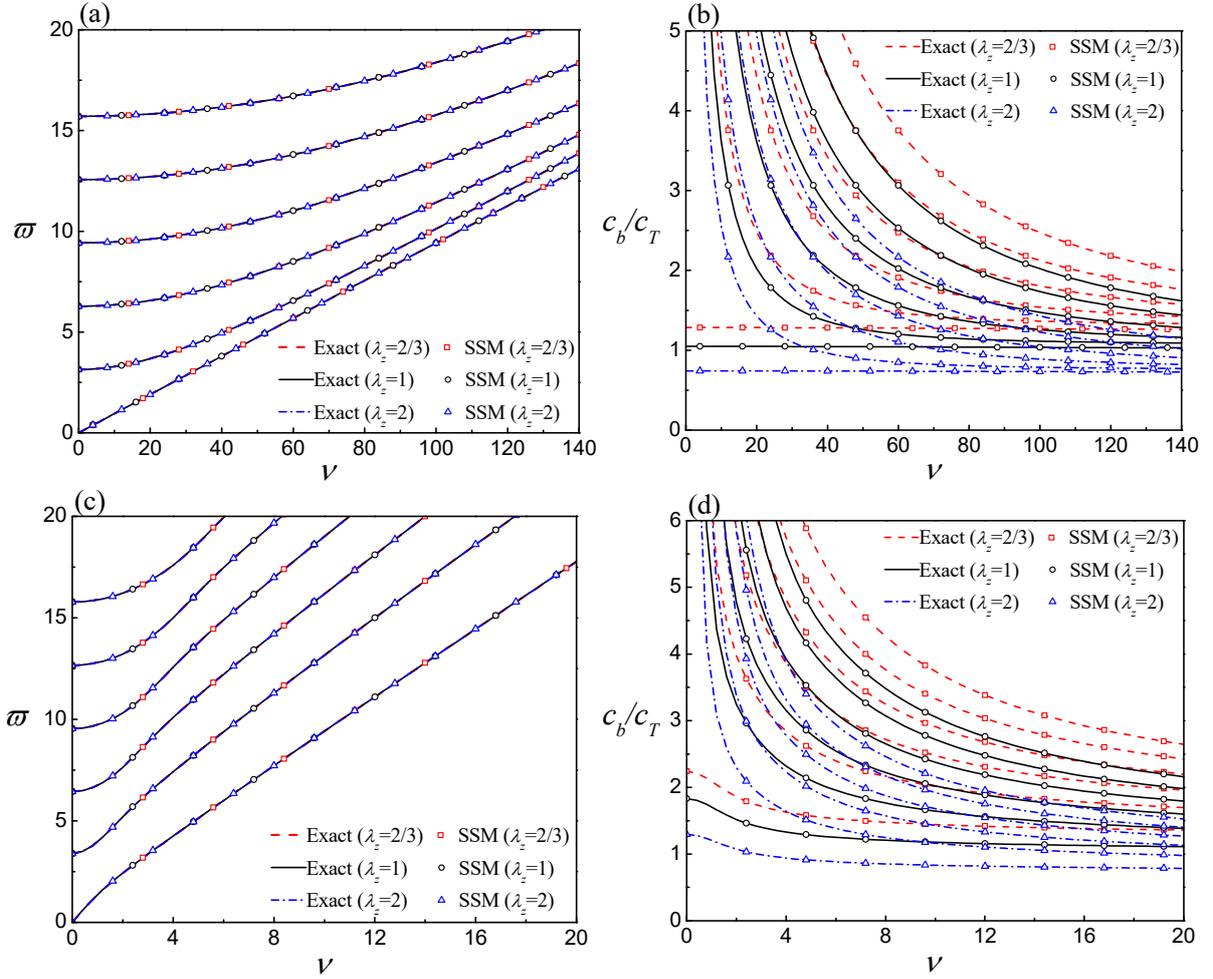

**Fig. 5.** Comparison of the frequency spectra (a, c) and phase velocity spectra (b, d) of the first six SH wave modes obtained by the exact solutions and the SSM for thin (a, b) and thick (c, d) EA tubes at different pre-stretches.

As we have already mentioned, the deformation of the pre-stretched EA tube will be homogeneous when no electric voltage is applied to the EA tube. In this situation, we can obtain exact dispersion relations for both SH and Lamb waves, which are given in Appendix C. Therefore, the SSM may be further verified by making a comparison of the numerical results with these exact solutions. For the neo-Hookean material and three different pre-stretches $\lambda_z = 2$, 1 and 2/3, the dispersion relation $(60)_1$ of the SH waves obtained by the SSM and the exact one given by Eq. (C.2) are compared in Fig. 5, while the comparison for the Lamb waves, i.e. Eq. $(60)_2$ versus Eq. (C.3), is displayed in Fig. 6. The first six branches of the frequency spectra ($\varpi - \nu$ curves) for the SH waves are plotted in Figs. 5(a) and 5(c) for the thin ($\eta = 1/1.1$) and thick ($\eta = 1/5$) tubes, respectively, while those for the Lamb waves are displayed in Figs. 6(a) and 6(c). At the same time, we also plot the first six branches of the phase velocity spectra ($c_b/c_T - \nu$ curves) for the SH and Lamb waves in Figs. 5(b), 5(d) and 6(b), 6(d), respectively. In Figs. 5 and 6, the lines correspond to the exact solutions while the markers to the SSM. Here, the number of the discretized sub-layers is taken to be 80 and 120, respectively, for the thin and thick EA tubes. As shown in Figs. 5 and 6, the SSM results agree very well with the exact solutions in the entire angular wave number range for both the thin and thick EA tubes. This excellent agreement again manifests the validation of the SSM.

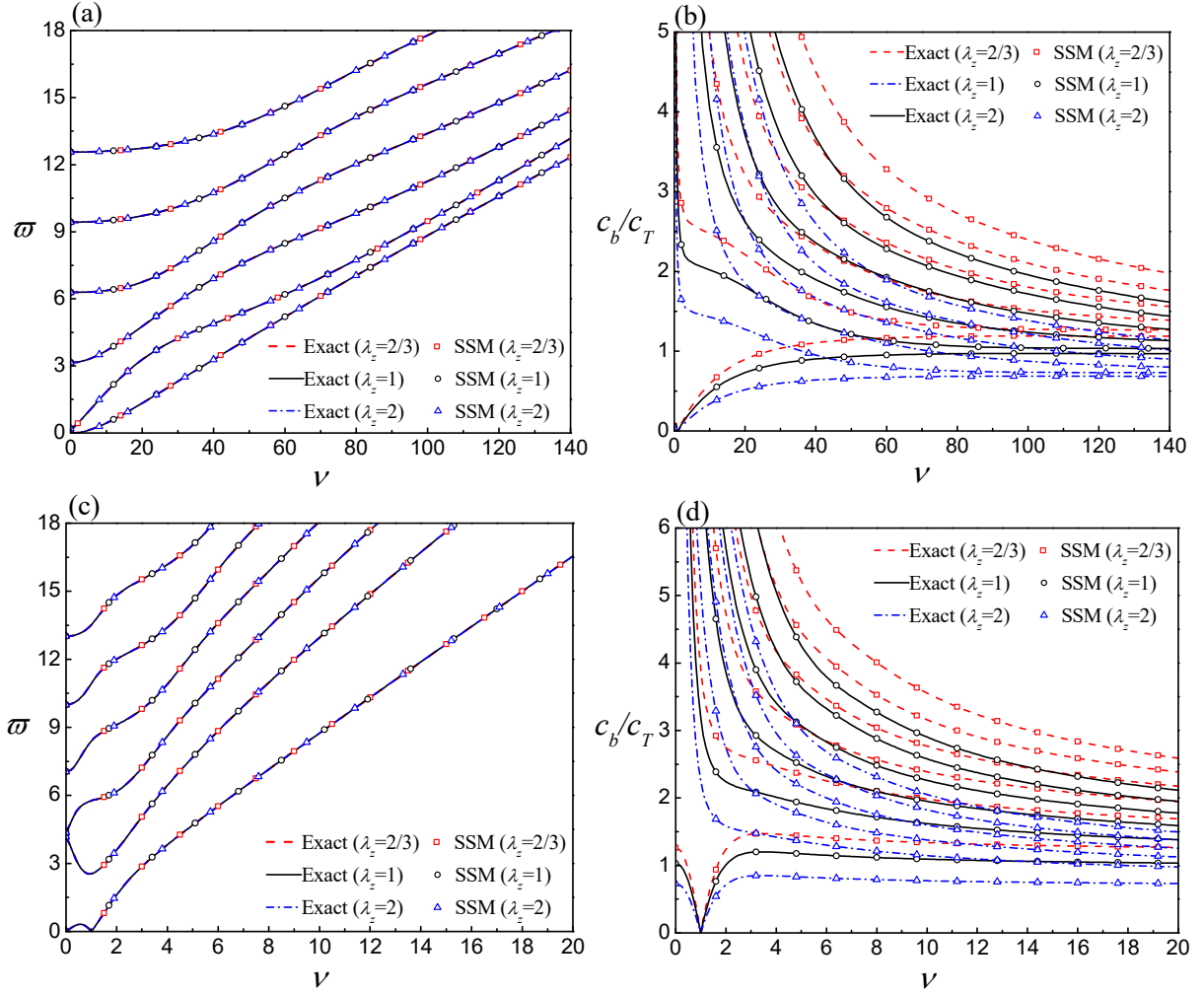

**Fig. 6.** Comparison of the frequency spectra (a, c) and phase velocity spectra (b, d) of the first six Lamb wave modes obtained by the exact solutions and the SSM for thin (a, b) and thick (c, d) EA tubes at different pre-stretches. (The frequency and the phase velocity of the Lamb waves become zero at $v = 1$, which means that there is no wave propagation in the pre-stretched hyperelastic tube when the wavelength on the outer surface of the EA tube is equal to $2\pi b$, the outer circumference.)

It is also seen from Figs. 5 and 6 that, for neo-Hookean materials, the frequency spectra for the SH and Lamb waves are not affected by the pre-stretches no matter what the thickness of the tube is, whereas the phase velocity of the circumferential waves traveling along the outer surface decreases as the pre-stretch $\lambda_z$ increases for a given excitation frequency $\varpi$. These phenomena are exactly in agreement with the analytical predictions given in Appendix C. In addition, we can also observe from Figs. 5 and 6 that, at high frequencies, the first branch curve of the frequency spectra for all cases is almost a straight line, indicating that the first mode is nondispersive. In fact, the first modes of the SH and Lamb waves at a large angular wave number asymptotically approach the modified shear wave velocity and the modified Rayleigh surface wave velocity, respectively, in the pre-stretched hyperelastic tube.

Interestingly, it is observed from Fig. 6 that, the frequency and the phase velocity of the Lamb waves become zero at $\nu = 1$, which means that there is no wave propagation in the pre-stretched hyperelastic tube when the wavelength on the outer surface of the EA tube is equal to $2\pi b$, the outer circumference. However, the frequency spectra and the phase velocity spectra for the Lamb waves in the EA tube still have branches at wave numbers ranging from $\nu = 0$ to $\nu = 1$, which is peculiar to the tube because of the multiple reflections between the inner and outer surfaces, as already noticed and discussed in Liu and Qu (1998).

In summary, the present SSM is extremely effective for studying the circumferential waves in a deformed EA tube with inhomogeneous biasing fields. In the following calculations, when the SSM is employed, the thin ($\eta = 1/1.1$) and thick ($\eta = 1/5$) EA tubes are divided into 80 and 120 equally thick layers, respectively, for which the results can be considered to be highly accurate.

*6.3. SH waves*

In this section, the SSM will be utilized to obtain the frequency spectra and the phase velocity spectra for SH waves propagating in EA tubes subjected to an axial pre-stretch and a radial electric voltage. The guided SH wave technique provides a reliable and efficient route to locating and sizing the axial defects or cracks in tubes (Zhao and Rose, 2004; Luo et al., 2005). Before developing a self-sensing EA actuator based on guided SH waves under biasing fields, the effects of the biasing fields on the wave propagation characteristics must be also understood and explored for both thin and thick tubes. As previously mentioned, there is a critical value of the electric voltage beyond which there exists no solution for the axisymmetric deformation, and the critical voltage depends on the geometry of the tube and the axial pre-stretch.

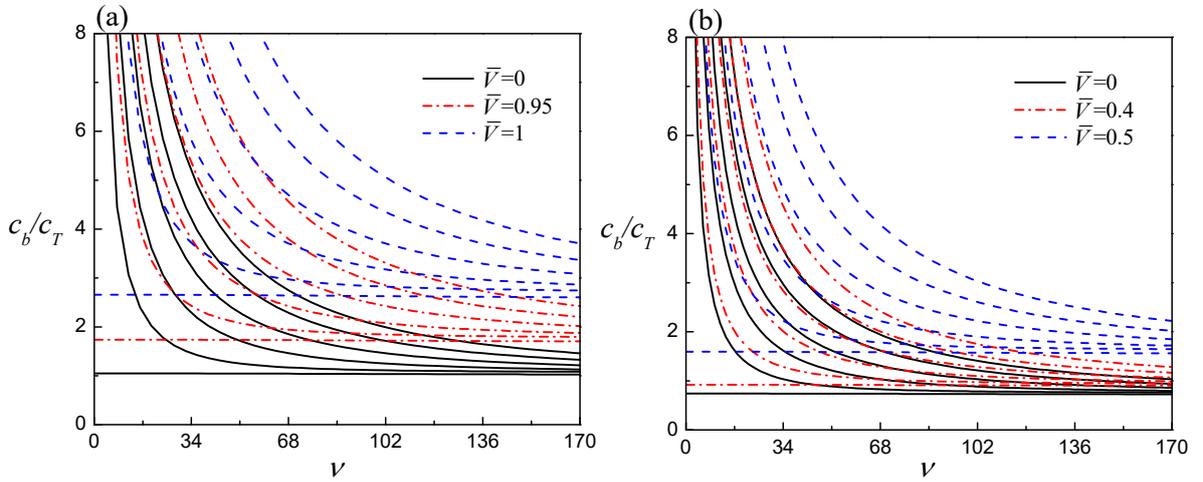

**Fig. 7.** Phase velocity spectra (with first six branches) of SH waves for a thin EA tube: (a) $\lambda_z = 1$; (b) $\lambda_z = 2$.

By numerical calculations, we have obtained the first six branches of the frequency spectra of the SH waves for both thin and thick tubes at three pre-stretches $\lambda_z = 2$, 1 and 2/3 and different allowable electric voltages. The results indicate that although the radial electric voltage produces radially inhomogeneous biasing fields, the electric voltage and pre-stretch scarcely influence the frequency spectra of the SH waves for a given EA tube (i.e., when $\eta$ is fixed). Therefore, the frequency spectra are nearly the same as those in Figs. 5(a) and 5(c), and thus omitted here for the sake of brevity.

The first six branches of the phase velocity spectra for the SH waves are shown in Figs. 7 and 8 for both thin and thick EA tubes, respectively, at different values of the dimensionless electric voltage $\bar{V}$ and pre-stretch $\lambda_z$. In particular, the results for $\bar{V}=0$, corresponding to the purely mechanical pre-stretch case in Figs. 5 and 6, are included in Figs. 7 and 8 for comparison. It is found that, although the frequency spectra of the SH waves are independent of the electric voltage, the phase velocities for both thin and thick tubes increase with the electric voltage in the entire wave number range. Furthermore, we can also observe that, in all instances, the first branches of the phase velocity spectra all start from a finite value depending on the biasing fields, whereas the higher branches start from infinity but with a finite cutoff frequency independent of the biasing fields. They all asymptotically approach at large wave numbers the modified shear wave velocities in the EA tubes with different biasing fields. For the thin tube in Fig. 7, the first modes in all cases are almost nondispersive in the entire range. However, for the thick tube in Fig. 8, the first modes are dispersive at a small angular wave number although they tend to be nondispersive at a large wave number. These characteristics greatly resemble the SH waves in an isotropic elastic tube (Zhao and Rose, 2004).

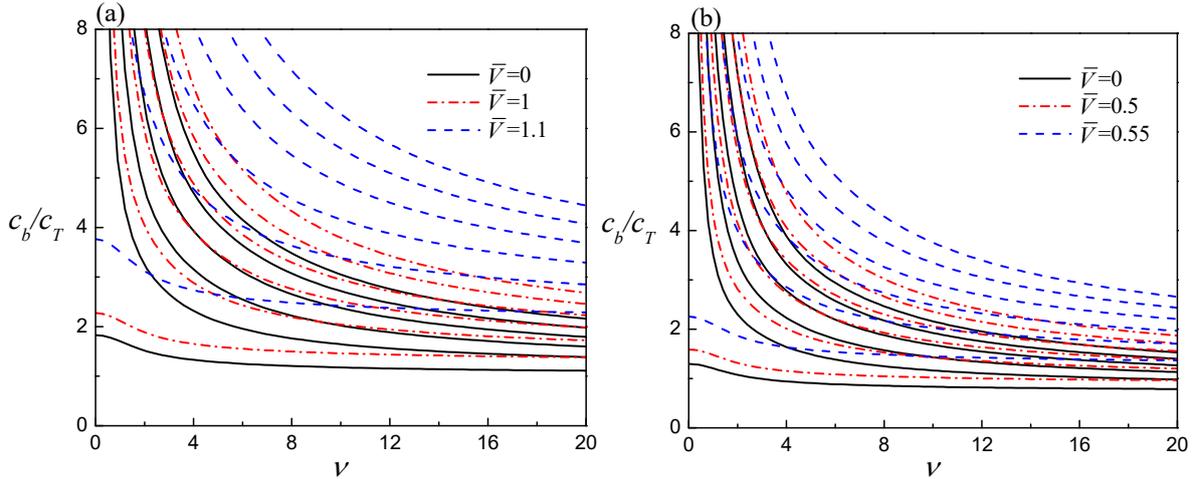

**Fig. 8.** Phase velocity spectra (with first six branches) of SH waves for a thick EA tube: (a) $\lambda_z=1$; (b) $\lambda_z=2$.

For the thin tube with $\lambda_z=1$, the displacements $\bar{U}_z(\xi)$ of the first two modes along the dimensionless radial coordinate $\xi$ are calculated at $\varpi=16$ (corresponding to $\nu=171.502$ and 162.482, respectively) and plotted in Figs. 9(a) and 9(b) for various radial electric voltages. In addition, Figs. 9(c) and 9(d) display the displacements $\bar{U}_z(\xi)$ of the first two modes in the thick tube with $\lambda_z=2$ also at $\varpi=16$ (corresponding to $\nu=17.861$ and 13.103, respectively). Generally speaking, a higher wave mode has a smaller angular wave number for a given circular frequency. It is worth mentioning that, the displacement amplitude is normalized within ±1 by its absolute maximum value for each electric voltage. Also, the starting point of the dimensionless radial coordinate becomes different when the biasing fields change. As shown in Fig. 9, the mode shape is essentially independent of the electric voltage although the thickness of the EA tube reduces with the increase of the electric voltage. Besides, we also plot in Fig. 10 the first six displacement modes of the thick tube with $\lambda_z = 1$ and $\bar{V} = 1.1$ at $\varpi = 16$ (the corresponding six angular wave numbers are indicated in the figure). It is remarked here that, the first mode decreases monotonically along the radial direction from the outer surface to the inner surface, while other wave modes are oscillatory along $\xi$. The peak value location moves inward with the mode number increasing, which means that more wave energy is concentrated near the inner surface in the case of higher wave modes. Thus, in order to increase the signal-to-noise ratio, high-order mode SH waves can be selectively generated to detect the fatigue cracks or defects near the inner surface of the soft EA tube, an access to which is usually unavailable (Zhao and Rose, 2004). In particular, there are $n-1$ zero-crossing points in the mode profile of the $n$th mode. These observations are consistent with those reported by Chen (1973).

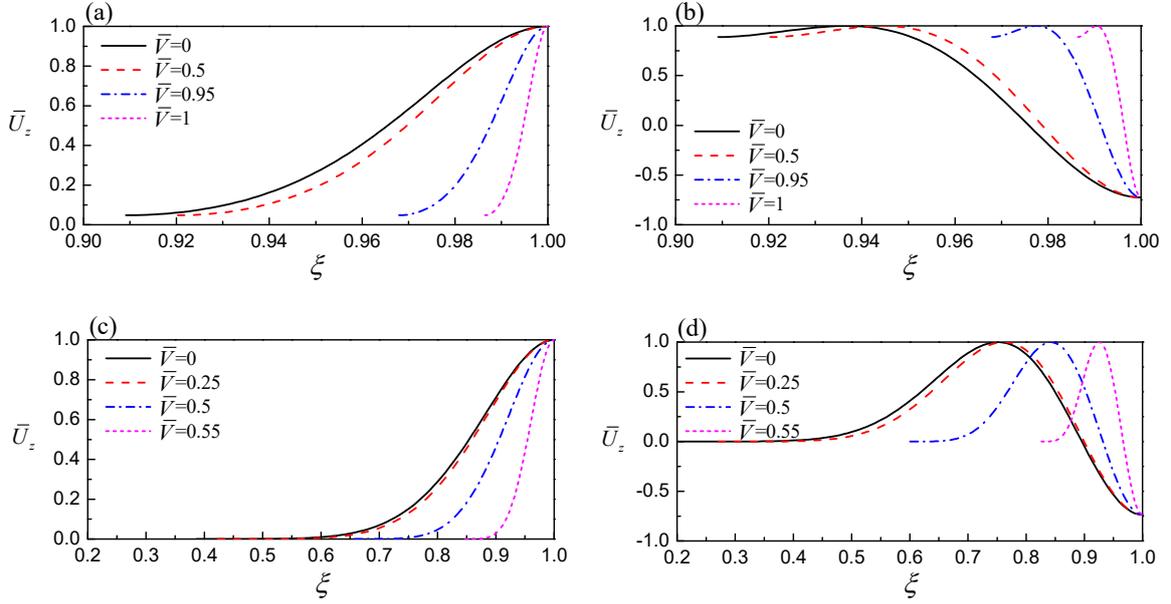

**Fig. 9.** Normalized displacement variation of SH waves for different values of $\bar{V}$ at $\varpi = 16$: (a, b) the first two modes for a thin EA tube at $\lambda_z = 1$; (c, d) the first two modes for a thick EA tube at $\lambda_z = 2$.

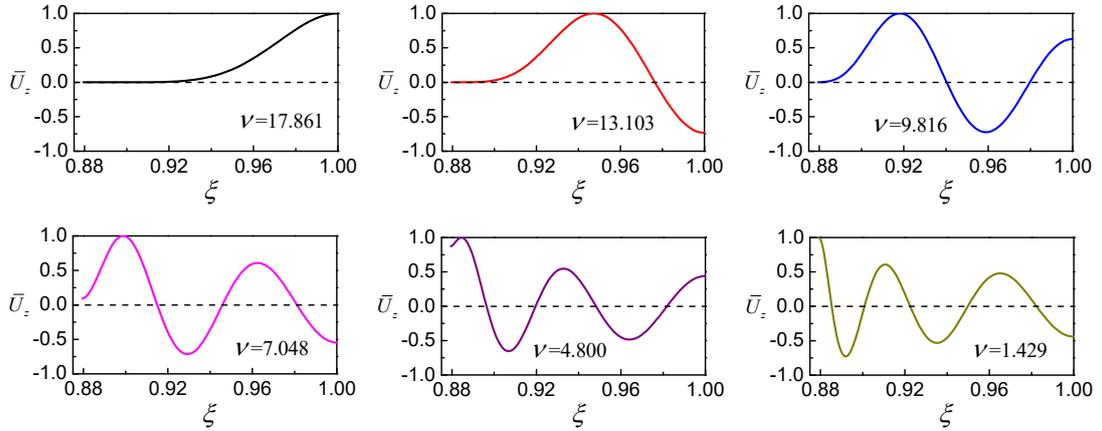

**Fig. 10.** Normalized displacement variation of the first six SH wave modes for a thick EA tube at $\lambda_z =1$, $\bar{V} =1.1$ and $\varpi =16$.

As shown in Figs. 7 and 8, the phase velocity along the outer surface increases with the radial electric voltage. In order to clearly display the dependence of the dimensionless phase velocity on the biasing electric field, the curves of the dimensionless phase velocity of the first mode versus the electric voltage are depicted in Fig. 11 for different pre-stretches at two representative values of the wave number $\nu =170$ and $\nu = 20$ for both thin and thick EA tubes. It can be seen that the phase velocity increases monotonically with and depends nonlinearly on the applied electric voltage. Specifically, this rise is small for a low electric voltage, while a huge phase velocity enhancement is observed when the electric voltage tends to the critical value $\bar{V}_c$. As described in Section 6.1, the EA tube expands rapidly when the electric voltage approaches the critical voltage. As a result, the phase velocity will undergo a drastic increase according to Eqs. (52)$_1$ and (54)$_2$ although the $\varpi-\nu$ relationship is not affected by the applied electric voltage. In addition, since the critical electric voltage for $\lambda_z = 2$ is lower than that for $\lambda_z =1$, the same level of the phase velocity rise exists at a lower electric voltage in the former case. That is to say, the tensile pre-stretch enhances the performance of the EA tube, which is in accordance with the findings by Shmuel (2015) on torsional motions of soft dielectric tubes subjected to electromechanical biasing fields. To conclude, the phase velocity of the guided SH waves varies nonlinearly with the electric voltage in an EA tube, which provides a possibility to develop a self-sensing EA actuator with actuating and sensing simultaneously. Specifically, by sensing the variations of the voltage-dependent phase velocity, the applied actuation voltage and hence the actuation strain or displacement may be precisely controlled.

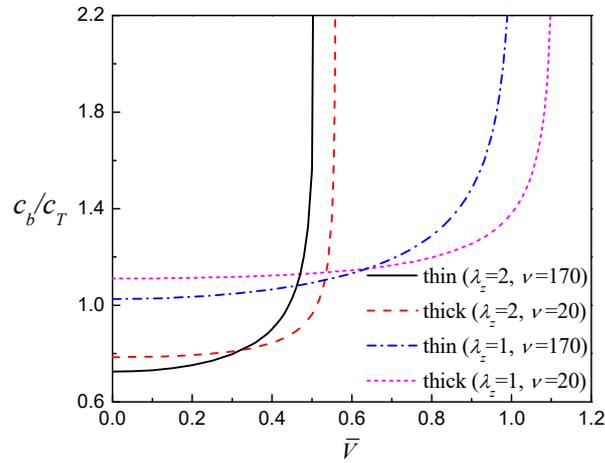

**Fig. 11.** Variations of the dimensionless phase velocity $c_b/c_T$ with $\bar{V}$ for both thin and thick tubes at different pre-stretches ($\lambda_z =1$ and $\lambda_z = 2$) and angular wave numbers ($\nu =170$ and $\nu = 20$).

*6.4. Lamb waves*

For radial defects or cracks which are often found to initiate from the inner surface of tubes, the guided Lamb waves can be used appropriately to detect and characterize them in a nondestructive manner (Liu and Qu, 1998; Valle et al., 2001). Thus, we now study the effects of the electromechanical biasing fields on the guided Lamb waves propagating in an EA tube. The geometry of the tubes and the biasing fields under consideration are assumed to be the same as those adopted in the previous subsection for the SH waves.

As pointed out by Valle et al. (2001) and Giurgiutiu (2008), due to the multi-mode character of guided waves travelling in waveguides and after the interaction of the guided waves with a structural defect or fatigue crack, the received signal generally contains more than one mode and the proportion of different modes depends on the mode conversion phenomenon at defects. Additionally, guided wave modes are generally dispersive, which means that the wave shape will change with the distance along the propagation path. The defect sensitivity of different wave modes in different frequency regions is a critical parameter that determines the best testing regime for different defect types. Consequently, it is significantly important to understand the dispersion feature of the guided Lamb waves before carrying out a nondestructive testing.

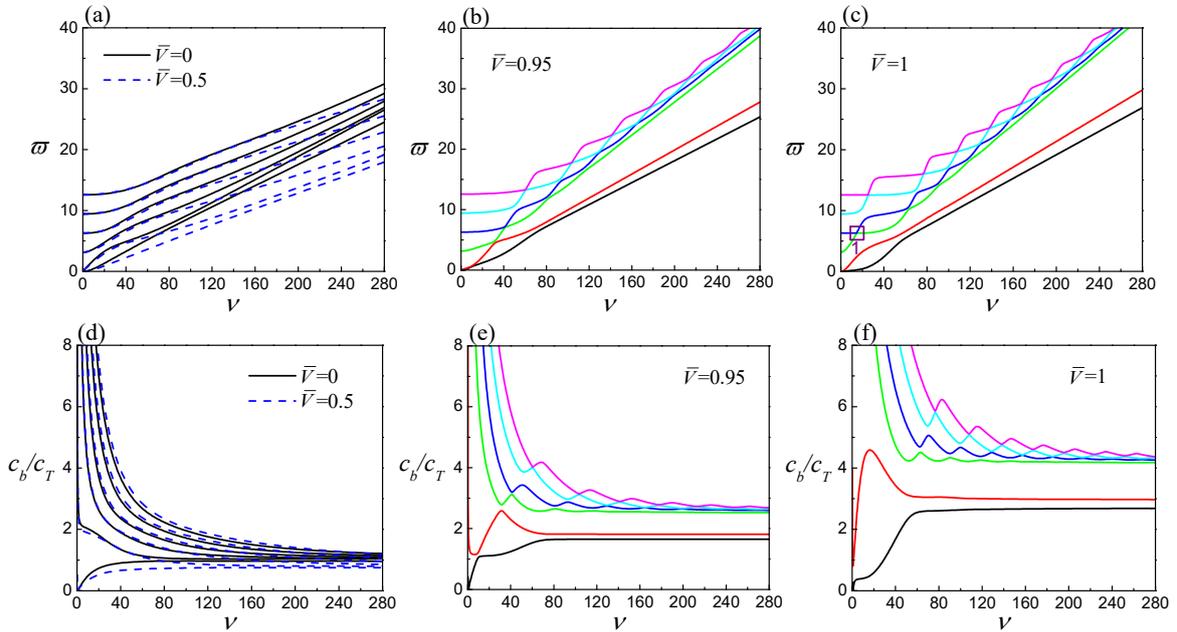

**Fig. 12.** Frequency spectra (a-c) and phase velocity spectra (d-f) of the first six Lamb wave modes for a thin EA tube at $\lambda_z = 1$ and different electric voltages. The rectangle labeled with 1 in (c) indicates a frequency veering region.

For a pre-stretch $\lambda_z = 1$ and different dimensionless electric voltages $\bar{V}$, the frequency spectra and the phase velocity spectra for the first six branches of the Lamb waves are presented in Fig. 12 for a thin EA tube. Unlike SH waves considered in the previous section, the applied electric voltage indeed shows a considerable influence on the frequency spectra as well as the phase velocity spectra for the Lamb waves. Specifically, for the considered thin EA tube in Fig. 12, it is seen that a low electric voltage corresponding to $\bar{V} = 0.5$ plays a significant role at a large angular wave number for high-order wave modes although it hardly affects the frequency spectra at a relatively small wave number. In addition, the lower-order modes are more sensitive to the applied electric voltage than the higher-order modes, especially at large angular wave numbers. In particular, a low electric voltage obviously decreases the frequency and the phase velocity of the first two wave modes in the entire angular wave number range.

As the applied electric voltage increases up to the critical value, the frequency spectra and the phase velocity spectra exhibit substantial changes. Similar to the low electric voltage case, the first two branches of the frequency spectra for high electric voltages ($\bar{V} = 0.95$ and 1 here) are also straight lines with different slopes at large wave numbers. This implies physically that the first two modes are nondispersive and correspond to modified Rayleigh-type surface waves on the outer and inner surfaces of the EA tube as described below. However, the frequencies and the phase velocities as well as the gaps between the first two modes for a given large angular wave number increase with the electric voltage. Furthermore, the variations of the normalized displacement amplitude $\bar{U}_n(\xi) = \sqrt{|U_r(\xi)|^2 + |U_\theta(\xi)|^2}/H_n$ of the first two modes at $\nu = 240$ are shown in Fig. 13 for four different electric voltages $\bar{V} = 0$, 0.5, 0.95 and 1. Here, $H_n$ is the maximum absolute displacement amplitude along $\xi$. Note that in our numerical calculation, the normalized amplitudes $\bar{U}_r(\xi)$ and $\bar{U}_\theta(\xi)$ of the radial and circumferential displacements are real and pure imaginary, respectively, indicating that they are always 90 degrees out of phase (Liu and Qu, 1998). It can be observed from Fig. 13 that the first two modes at large angular wave numbers behave like Rayleigh-type surface waves on the outer and inner surfaces of the deformed EA tube, respectively, for all admissible electric voltages. However, the decay rate of the wave modes at a low electric voltage ($\bar{V} = 0.5$) is much slower than that at a high electric voltage ($\bar{V} = 0.95$ or 1). Owing to the nondispersive characteristic, i.e., invariable group velocity at almost all frequencies, the modified Rayleigh-type surface wave modes under biasing fields can be utilized to locate and charaterize the radial defects or cracks on the outer or inner surface of the soft EA tube (Valle et al., 2001).

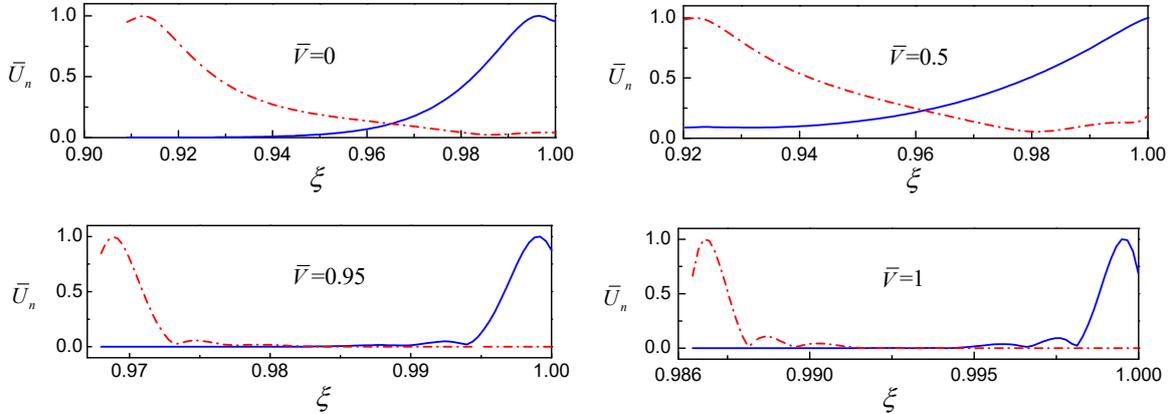

**Fig. 13.** Normalized displacement variation of the first two Lamb wave modes of a thin EA tube at $\lambda_z = 1$ and $\nu = 240$ for different electric voltages $\bar{V}$ (solid line: first mode; dash-dotted line: second mode).

Besides, we can also observe from Fig. 12 that, there is no strong mode coupling under the low electric voltage $\bar{V} = 0.5$. Nonetheless, for high electric voltages $\bar{V} = 0.95$ and 1, the higher-order modes for the frequency spectra and the phase velocity spectra may group together in pairs apparently, exhibiting an oscillatory behavior. In particular, there exists the special frequency veering phenomenon between the adjacent modes, which means that the two curves of the frequency spectra come close and do not cross but rather veer apart from each other with a high local curvature (Mace and Manconi, 2012). It should be pointed out that, close to the frequency veering region, the interaction between the two branches becomes substantial and the wave mode shapes change rapidly around the region, which usually results in significant flows of the mechanical energy between the two neighboring wave modes (Liu and Qu, 1998; Mace and Manconi, 2012; Zhu et al., 2013). For example, the frequency veering region between the third and the fourth branches for $\bar{V}=1$ is indicated by the rectangle in Fig. 12(c). In order to have a closer look, that region is enlarged and shown in Fig. 14.

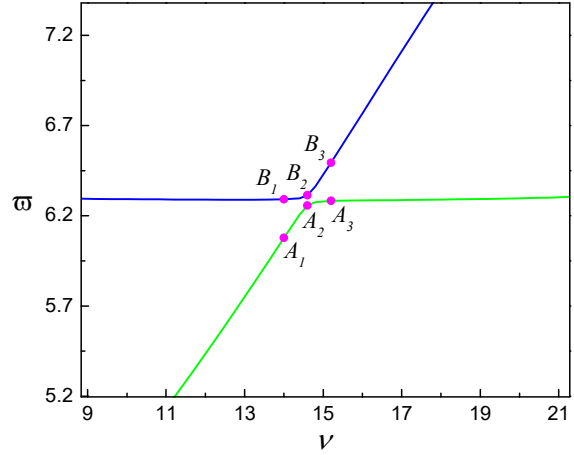

**Fig. 14.** Enlarged view of the frequency veering region indicated by the rectangle in Fig. 12(c).

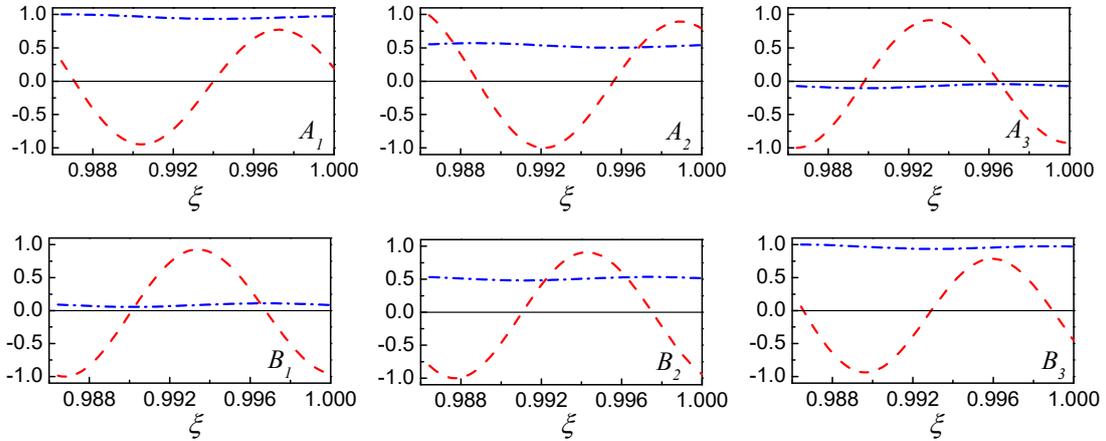

**Fig. 15.** Conversion of the wave mode shapes (dashed line: normalized circumferential displacement $\bar{U}_\theta(\xi)$; dash-dotted line: normalized radial displacement $\bar{U}_r(\xi)$).

The wave mode shapes corresponding to different marked points within the frequency veering region in Fig. 14 are depicted in Fig. 15. The dashed and dash-dotted curves denote the variations of the normalized circumferential and radial displacements $\bar{U}_\theta(\xi)$ and $\bar{U}_r(\xi)$, respectively, along the dimensionless radial coordinate $\xi$. Note that the mode shapes are normalized by the maximum absolute value among the two displacement components. It can be clearly seen from Fig. 15 that, at point $A_1$ on the third branch, the radial displacement (i.e., thickness-stretch) predominates and the circumferential displacement (i.e., thickness-shear) exhibits a nearly antisymmetric profile with respect to the mid-plane of the deformed EA tube, while at point $A_3$, the circumferential displacement prevails, almost symmetric about the mid-plane. On the contrary, the wave mode shape changes from $B_1$ to $B_3$ on the fourth branch in almost a reverse way. Consequently, a slight variation of the frequency can alter the wave mode shape significantly. As a result, precautions should be taken on the wave mode shape selection when deciding to use the frequency veering region for detecting and characterizing defects and cracks due to the conversion of the wave mode shapes.

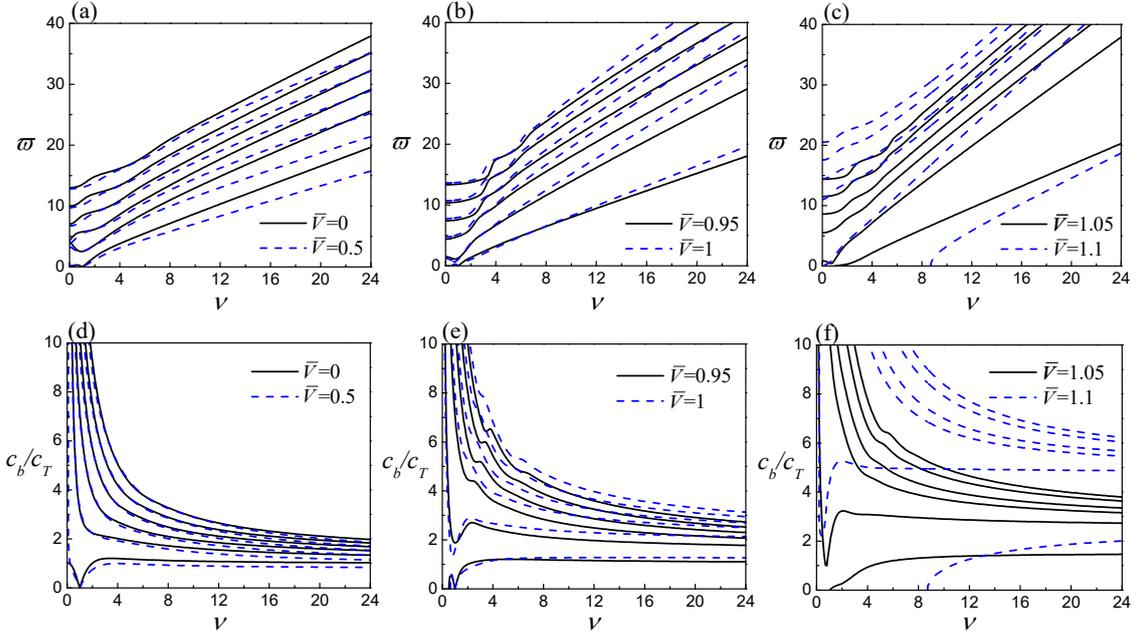

**Fig. 16.** Frequency spectra (a-c) and phase velocity spectra (d-f) of the first six Lamb wave modes for a thick EA tube at $\lambda_z = 1$ for different electric voltages.

For the thick EA tube with a pre-stretch $\lambda_z = 1$ and different dimensionless electric voltages $\bar{V}$, we also plot the frequency spectra and the phase velocity spectra for the first six branches of the guided Lamb waves in Fig. 16. It can be seen here that, there is no strong mode coupling phenomenon, even when the applied electric voltage is high. This is clearly distinct from that for the thin EA tube. In addition, the frequency and the phase velocity of all wave modes are only slightly reduced by the low electric voltage ($\bar{V} = 0.5$) in the entire angular wave number range. However, a high electric voltage increases the frequency and the phase velocity of the higher-order wave modes in the entire wave number range except for the second mode at small wave numbers. Another interesting phenomenon is the existence of a cutoff angular wave number in the first branch of the frequency spectra for a high electric voltage, i.e. the first branch intersects with the horizontal ordinate at a finite cutoff wave number. The cutoff wave number defines a critical wavelength beneath which there is no stable propagation of longer waves for the first mode. It is seen here that the cutoff wave number increases monotonically with the electric voltage. This phenomenon is pretty similar to that of the first antisymmetric mode of the generalized Rayleigh-Lamb waves in DE layers subjected to electromechanical biasing fields (Shmuel et al., 2012).

In order to clearly reveal the effects of the applied electric voltage on the Lamb wave propagation characteristics, the variations of the dimensionless phase velocity of the first three Lamb wave modes with the applied electric voltage are displayed in Fig. 17 at a representative frequency $\varpi = 15$ for both thin and thick EA tubes with $\lambda_z = 1$. Notably, the phase velocity no longer increases monotonically, which is quite different from the results of the SH waves. Specifically, as the applied electric voltage increases from zero to the critical value, the phase velocity will decrease to a minimum (at $\bar{V} = 0.5$, 0.45 and 0.44 for the thin EA tube and $\bar{V} = 0.7$, 0.54 and 0.42 for the thick EA tube for the first three modes, respectively), and then increase considerably. The reason for this phenomenon can be explained as follows. According to Eqs. (52)$_1$ and (54)$_2$, the phase velocity along the outer surface of the deformed EA tube is proportional to the outer radius $b$ and inversely proportional to the angular wave number $\nu$ for a given frequency $\varpi$. When the applied electric voltage is low, which produces relatively small biasing fields and alters the stiffness of the EA tube only a little, the EA tube expands extremely slightly and its thickness decreases relatively remarkably due to the small variation of $\lambda_a$ and the large increase in the radius ratio $\bar{\eta}$ of the deformed EA tube, as can be seen from Figs. 3(a) and(c). As a result, the wave number $\nu$ at a fixed frequency increases, as shown in Figs. 12(a) and 16(a). This behavior is similar to that in the classical isotropic and linear elastic case (Liu and Qu, 1998). Since the increase in the wave number $\nu$ surpasses that in the outer radius $b$, the phase velocity is reduced by the low electric voltage. For a high electric voltage, relatively large inhomogeneous biasing fields are generated and the stiffening effects become significant. Thus, the wave number $\nu$ corresponding to a given frequency decreases, which can be observed from Figs. 12(b)-(c) and 16(b)-(c), although the shape factor $\bar{\eta}$ still increases in this case. In addition, the outer radius $b$ increases sharply as shown in Fig. 3(a). Consequently, the phase velocity increases remarkably, especially when the applied electric voltage tends to the critical value. In brief, the nonlinear dependence of the phase velocity on the applied electric voltage is a consequence of the competition between the changes in the shape factor $\bar{\eta}$ and the outer radius $b$ of the deformed EA tube. Analogous to the SH waves, a self-sensing EA tube actuator can be developed, which operates based on the variation of the phase velocity of guided Lamb waves when an electric voltage is applied to the EA tube. Also interestingly, this phenomenon may be exploited to manipulate the delay time of the Lamb waves in delay lines by properly adjusting the applied electric voltage. As already mentioned previously, the changes in the phase velocity and the mode shapes of the guided Lamb waves can be utilized in the ultrasonic nondestructive on-line monitoring or structural health monitoring (SHM) of the EA tube actuators.

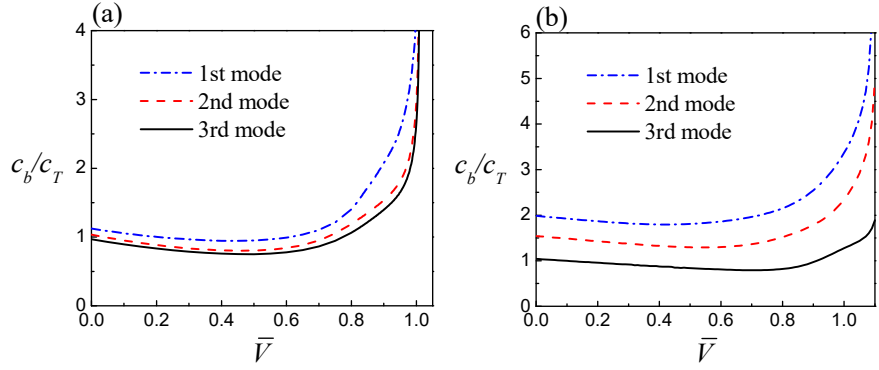

**Fig. 17.** Variations of the dimensionless phase velocity $c_b/c_T$ for the first three Lamb wave modes with $\bar{V}$ at $\lambda_z = 1$ and $\varpi = 15$: (a) thin EA tube; (b) thick EA tube.

For both thin and thick EA tubes, numerical calculations have also been conducted for different pre-stretches. The results are qualitatively similar to those in Figs. 12 and 16 and thus not reported here.

## 7. Conclusions

In this study, the guided circumferential waves propagating in soft EA tubes under electrostatically induced inhomogeneous biasing fields are analyzed for ultrasonic non-destructive structural health monitoring (SHM) and self-sensing purposes. First, the axisymmetric deformation of the soft EA tube with electrodes on its cylindrical surfaces subjected to an axial pre-stretch (or force) and a radial electric voltage is addressed by using a general strain energy function. In order to treat the radially inhomogeneous biasing fields, a method based on the state-space formalism and an approximate laminate model is developed. It should be noted here that another method based on the state-space formalism but making use of the Taylor's expansion may also be used (Chen, 2001). The dispersion relations of the SH and Lamb waves are then established efficiently and accurately. In particular, explicit expressions of the radially inhomogeneous biasing fields in a soft EA tube subjected to an axial pre-stretch and a radial electric voltage are derived for neo-Hookean ideal dielectric materials. It is found that, for a fixed pre-stretch, the EA tube expands, its thickness decreases, and the surface charge accumulates on the electrodes in response to the radial electric voltage in a nonlinear way. In addition, the degree of the radial inhomogeneity of the physical quantities increases with the increasing radial electric voltage. Afterwards, the efficiency of the proposed analysis method is verified by checking its convergence and accuracy. Numerical examples are finally presented to highlight the effects of the axial pre-stretch, the radial electric voltage, and the geometrical parameters on the SH and Lamb wave propagation characteristics.

For SH waves, the radial electric voltage and the pre-stretch scarcely influence the frequency spectra for a given soft neo-Hookean EA tube. On the other hand, the phase velocity along the outer surface of the deformed EA tube increases monotonically with the electric voltage, especially when the electric voltage approaches the critical value. Besides, the mode shape is essentially independent of the radial electric voltage for a fixed pre-stretch and the peak value location of the displacement modes moves inward with the mode number increasing. The behavior of the guided Lamb waves under radially inhomogeneous biasing fields is more complex than that of the SH waves and depends strongly on the mode, the wave number range, the applied biasing fields and the geometrical parameters. For Lamb waves, a frequency veering occurs for the thin EA tube when a high electric voltage is applied. In such a case, a slight variation of the frequency can alter the wave mode shape significantly. In addition, the first branch of the frequency spectra of the thick EA tube subjected to a high electric voltage emerges only after a cutoff angular wave number, which defines the critical wavelength beneath which there is no stable propagation of longer waves of the first mode. Unlike the SH waves, the phase velocity of the Lamb waves first decreases to a minimum and then increases considerably when the radial electric voltage increases from zero and up to the critical value. This complex variation is a consequence of the competition between the changes in the shape factor and the outer radius of the deformed EA tube. This work provides a theoretical reference for 1.) on-line ultrasonic non-destructive structural health monitoring (SHM) to detect and characterize interior defects and fatigue cracks in soft EA tube actuators utilizing the guided circumferential wave techniques, and 2.) self-sensing, control and adjusting of the actual operating electric voltage of the soft EA tube actuators based on the measured guided circumferential wave propagation characteristics. Indeed, these two interesting topics demand further research works.

The soft EA tube investigated in this analysis is assumed to be incompressible, and its material nonlinearity is characterized only by the neo-Hookean ideal dielectric model. Consequently, further analysis on the influences of the material compressibility and other nonlinear material models such as Yeoh model, Gent model, Arruda-Boyce model (which can be used to describe the strain-stiffening effects) on guided circumferential waves propagating in soft EA tubes is required. The analysis along this line is now under way.

Finally, we emphasize here that the proposed analysis method combining the state-space formalism with an approximate laminate technique can be conveniently applied to other types of guided waves such as waves propagating along the axis of the soft EA tube under radially inhomogeneous biasing fields and to different geometrical configurations such as EA spherical shells. In addition, wave propagation and vibration in laminated soft EA structures and functionally graded nonlinear EA structures subjected to electromechanical biasing fields can be analyzed in a similar way.

**Acknowledgments**

The work was supported by the National Natural Science Foundation of China (Nos. 11272281, 11321202, 11532001, and 11202182), the German Research Foundation (DFG, Project-No: ZH 15/20-1), and the China Scholarship Council (CSC). Partial support from the Fundamental Research Funds for the Central Universities (No. 2016XZZX001-05) is also acknowledged.

# Appendix A. Nomenclature and list of symbols

| | Nomenclature glossary | | |
|---|---|---|---|
| $\mathcal{B}_r, \mathcal{B}, \mathcal{B}_t$ | Regions of undeformed, initial, and current configurations | $\partial \mathcal{B}_r, \partial \mathcal{B}, \partial \mathcal{B}_t$ | Boundaries of $\mathcal{B}_r$, $\mathcal{B}$, and $\mathcal{B}_t$ |
| $t_0, t$ | Different time instants | $d\mathbf{X}, d\mathbf{x}$ | Line elements of $\mathcal{B}_r$ and $\mathcal{B}_t$ |
| $dA, da_t$ | Surface elements of $\mathcal{B}_r$ and $\mathcal{B}_t$ | $dV, dv$ | Volume elements of $\mathcal{B}_r$ and $\mathcal{B}_t$ |
| $\mathbf{X}, \mathbf{x}, \mathbf{y}$ | Position vectors of $\mathcal{B}_r$, $\mathcal{B}$, and $\mathcal{B}_t$ | $\mathbf{b}, \mathbf{C}$ | Left/right Cauchy-Green strain tensors |
| $\mathbf{N}, \mathbf{n}, \mathbf{n}_t$ | Outward unit normal vectors of $\mathcal{B}_r$, $\mathcal{B}$, and $\mathcal{B}_t$ | $\mathbf{F}, J$ | Deformation gradient tensor and its determinant |
| $\mathbf{E}, \mathbf{D}$ | Electric field and electric displacement vectors in $\mathcal{B}_t$ | $\mathcal{E}, \mathcal{D}$ | Electric field and electric displacement vectors in $\mathcal{B}_r$ |
| $\sigma_f$ | Free surface charge density on $\partial \mathcal{B}_t$ | $\rho$ | Material mass density |
| $p$ | Lagrange multiplier | $\boldsymbol{\tau}, \mathbf{T}$ | Total Cauchy/nominal stress tensors |
| $\Omega(\mathbf{F}, \mathcal{D})$ | Total energy density function per unit reference volume | $\mathbf{t}^a$ | Applied mechanical traction vector per unit area of $\partial \mathcal{B}_t$ |
| $I_m$ | Invariants of $\Omega$ for an incompressible isotropic electroelastic material | $\Omega_m$ | Derivative of $\Omega$ with respect to invariants $I_m$ |
| $\mathbf{u}, u_m$ | Incremental displacement vector and its components | $\dot{\mathcal{E}}_0, \dot{\mathcal{D}}_0, \dot{\mathbf{T}}_0$ | Push-forward versions of Langrangian increments |
| $\dot{p}$ | Incremental Lagrange multiplier | $\dot{\mathcal{E}}_{0m}, \dot{\mathcal{D}}_{0m}, \dot{T}_{0mn}$ | Components of $\dot{\mathcal{E}}_0, \dot{\mathcal{D}}_0, \dot{\mathbf{T}}_0$ |
| $\mathbf{H}$ | Push-forward incremental displacement gradient tensor with respect to $\mathcal{B}$ | $\dot{\sigma}_{F0}, \dot{\mathbf{t}}_0^A$ | Incremental surface charge density and mechanical traction vector on $\partial \mathcal{B}$ |
| $\mathcal{A}, \mathcal{M}, \mathcal{R}$ | Referential electroelastic moduli tensors | $\mathcal{A}_0, \mathcal{M}_0, \mathcal{R}_0$ | Instantaneous electroelastic moduli tensors |
| $\mathcal{A}_{ijkl}, \mathcal{M}_{ijk}, \mathcal{R}_{ij}$ | Components of $\mathcal{A}, \mathcal{M}, \mathcal{R}$ | $\mathcal{A}_{0ijkl}, \mathcal{M}_{0ijk}, \mathcal{R}_{0ij}$ | Components of $\mathcal{A}_0, \mathcal{M}_0, \mathcal{R}_0$ |
| $A, B, L, a, b, l$ | Inner and outer radii, and length of undeformed/deformed EA tube | $R, \Theta, Z, r, \theta, z$ | Cylindrical coordinates in undeformed/deformed configuration |
| $\lambda_r, \lambda_z$ | Circumferential and axial stretches | $V$ | Initial radial electric voltage difference |
| $\lambda_a, \lambda_b$ | Circumferential stretches of inner/outer surfaces | $H, h$ | Thicknesses of undeformed/deformed EA tube |
| $\eta, \bar{\eta}$ | Ratios of inner radius to outer radius of undeformed/deformed EA tube | $\mathcal{D}_r, D_r$ | Initial radial electric displacements in $\mathcal{B}_r$ and $\mathcal{B}$ |
| $E_r$ | Initial radial electric field in EA tube | $\phi$ | Initial electrostatic potential in EA tube |
| $Q(a), Q(b)$ | Free surface charges on inner/outer surfaces of EA tube | $\Omega^*(\lambda_r, \lambda_z, I_4)$ | Reduced energy density function for axisymmetric deformations |
| $\sigma_{fa}, \sigma_{fb}$ | Free surface charge densities on inner/outer surfaces of EA tube in $\mathcal{B}$ | $\tau_{rr}, \tau_{\theta\theta}, \tau_{zz}$ | Initial normal stress components of EA tube in $\mathcal{B}$ |
| $e_{ip}, \varepsilon_{ij}, c_{mn}$ | Material parameters defined in Eq. (41) | $\dot{\phi}$ | Incremental electric potential in EA tube |
| $N$ | Resultant axial force on each end of deformed EA tube | $\mathbf{I}$ | Second-order identity tensor |
| $\mathbf{Y}, \mathbf{M}$ | Incremental state vector and $8 \times 8$ system matrix | $\mathbf{M}_{ij}$ | Four partitioned $4 \times 4$ sub-matrices of $\mathbf{M}$ |
| $\mathbf{Y}_1, \mathbf{M}_1, \mathbf{V}_1, \bar{\mathbf{M}}_1$ | Incremental state vectors and $2 \times 2$ system matrices for SH wave | $\mathbf{Y}_2, \mathbf{M}_2, \mathbf{V}_2, \bar{\mathbf{M}}_2$ | Incremental state vectors and $6 \times 6$ system matrices for Lamb wave |
| $n_1, n_2, q_m$ | Material parameters defined in system matrix $\mathbf{M}$ | $U_r, U_\theta, U_z,$ $\Phi, \Sigma_{0rr}, \Sigma_{0r\theta},$ $\Sigma_{0rz}, \Delta_{0r}$ | Modal distribution of incremental fields along radial coordinate |
| $\xi$ | Dimensionless radial coordinate in deformed EA tube | $\mu, c_T$ | Shear modulus/wave velocity of EA tube in absence of an electric field |
| $\varepsilon$ | Dielectric constant of ideal dielectric material | $k_r, \nu$ | Linear/angular wave number of SH and Lamb waves in deformed EA tube |
| $c_r, \alpha$ | Linear/angular phase velocity of SH and Lamb waves in deformed EA tube | $\omega, \varpi$ | Circular/dimensionless frequency of SH and Lamb waves |
| $c_b$ | Phase velocity of SH and Lamb waves at outer surface of deformed EA tube | $s, \beta_i$ | Dimensionless quantities defined in Eq. (52) |
| $n$ | Divided sub-layer number of EA tube | $\mathbf{K}_k (k=1,2)$ | Global transfer matrix for SH/Lamb waves |
| $\xi_{j0}, \xi_{j1}, \xi_{jm}$ | Dimensionless radial coordinate at inner/outer/middle surfaces of $j$th layer | $\mathbf{V}_k^0, \mathbf{V}_k^1$ | Incremental state vectors at inner/outer surfaces of deformed EA tube |
| $\bar{Q}, \bar{V}, \bar{N}$ | Dimensionless surface charge/electric potential difference/resultant axial force | $\bar{p}, \bar{\tau}_{rr}, \bar{\tau}_{\theta\theta},$ $\bar{\tau}_{zz}, \bar{D}_r$ | Dimensionless Lagrange multiplier/normal stresses/radial electric displacement |
| $\bar{V}_c$ | Critical radial electric voltage difference | $\bar{U}_r, \bar{U}_\theta,$ $\bar{U}_z, \bar{U}_n$ | Normalized displacement amplitudes of circumferential waves |
| $H_n$ | Maximum absolute displacement amplitude of Lamb waves | $\varphi, \psi$ | Scalar displacement functions of $u_r$ and $u_\theta$ |
| $\nabla^2$ | Two-dimensional Laplace operator | $\mathbf{S}$ | Coefficient matrix defined in Eq. (C.4) |
| $J_\nu, Y_\nu$ | Bessel function of first/second kind of order $\nu$ | $\delta, \gamma_1, \gamma_2$ | Material parameters defined in the dispersion equations (C.2) and (C.3) |

**Appendix B. Non-zero components of the instantaneous electroelastic moduli tensors**

According to the formulations by Dorfmann and Ogden (2010), we can obtain the following non-zero components of the instantaneous electroelastic moduli tensors

$$\mathcal{A}_{01111} = \frac{2}{\lambda_r^4 \lambda_z^4} \left\{ 2\Omega_{11} + \lambda_r^2 \lambda_z^2 \left[ \Omega_1 + \Omega_2(\lambda_r^2 + \lambda_z^2) \right] + 2(\lambda_r^2 + \lambda_z^2)\left[ 2\Omega_{12} + \Omega_{22}(\lambda_r^2 + \lambda_z^2) \right] + 2D_r^4 \left( 4\Omega_{66} + 4\Omega_{56}\lambda_r^2\lambda_z^2 + \Omega_{55}\lambda_r^4\lambda_z^4 \right) \right.$$
$$\left. + D_r^2 \left[ 8\Omega_{16} + \lambda_r^2\lambda_z^2 \left( 4\Omega_{15} + 6\Omega_6 + \Omega_5\lambda_r^2\lambda_z^2 \right) + 4(\lambda_r^2 + \lambda_z^2)(2\Omega_{26} + \Omega_{25}\lambda_r^2\lambda_z^2) \right] \right\},$$

$$\mathcal{A}_{01122} = \frac{4}{\lambda_r^2 \lambda_z^4} \left\{ (\Omega_{11} + \Omega_2)\lambda_r^2\lambda_z^2 + \Omega_{22}\left( \lambda_z^2 + \lambda_r^4\lambda_z^4 + \lambda_r^2(1+\lambda_z^6) \right) + \Omega_{12}\left( 1 + \lambda_r^4\lambda_z^2 + 2\lambda_r^2\lambda_z^4 \right) \right.$$
$$\left. + D_r^2 \left[ 2\Omega_{26}\left(1 + \lambda_r^2\lambda_z^4\right) + \lambda_r^2\lambda_z^2 \left( 2\Omega_{16} + \Omega_{25}(1+\lambda_r^2\lambda_z^4) + \Omega_{15}\lambda_r^2\lambda_z^2 \right) \right] \right\},$$

$$\mathcal{A}_{01133} = \frac{4}{\lambda_r^4 \lambda_z^2} \left\{ (\Omega_{11} + \Omega_2)\lambda_r^2\lambda_z^2 + \Omega_{22}\left( \lambda_r^2 + \lambda_r^4\lambda_z^4 + \lambda_z^2(1+\lambda_r^6) \right) + \Omega_{12}\left( 1 + \lambda_r^2\lambda_z^4 + 2\lambda_r^4\lambda_z^2 \right) \right.$$
$$\left. + D_r^2 \left[ 2\Omega_{26}\left(1 + \lambda_r^4\lambda_z^2\right) + \lambda_r^2\lambda_z^2 \left( 2\Omega_{16} + \Omega_{25}(1+\lambda_r^4\lambda_z^2) + \Omega_{15}\lambda_r^2\lambda_z^2 \right) \right] \right\},$$

$$\mathcal{A}_{02222} = \frac{2}{\lambda_z^4} \left\{ 2\Omega_{22}\left(1 + \lambda_r^2\lambda_z^4\right)^2 + \lambda_z^2 \left[ \Omega_2\left(1 + \lambda_r^2\lambda_z^4\right) + \lambda_r^2\lambda_z^2\left( \Omega_1 + 2\Omega_{11}\lambda_r^2 \right) + 4\lambda_r^2\Omega_{12}\left(1 + \lambda_r^2\lambda_z^4\right) \right] \right\},$$

$$\mathcal{A}_{02233} = \frac{4}{\lambda_r^2 \lambda_z^2} \left\{ (\Omega_{11} + \Omega_2)\lambda_r^4\lambda_z^4 + \lambda_r^2\lambda_z^2\Omega_{12}\left(2 + \lambda_r^4\lambda_z^2 + \lambda_r^2\lambda_z^4\right) + \Omega_{22}\left( 1 + \lambda_r^4\lambda_z^2 + \lambda_r^2\lambda_z^4 + \lambda_z^6\lambda_r^6 \right) \right\},$$

$$\mathcal{A}_{03333} = \frac{2}{\lambda_r^4} \left\{ 2\Omega_{22}\left(1 + \lambda_r^4\lambda_z^2\right)^2 + \lambda_r^2 \left[ \Omega_2\left(1 + \lambda_r^4\lambda_z^2\right) + \lambda_r^2\lambda_z^2\left( \Omega_1 + 2\Omega_{11}\lambda_z^2 \right) + 4\lambda_z^2\Omega_{12}\left(1 + \lambda_r^4\lambda_z^2\right) \right] \right\}$$

$$\mathcal{A}_{01212} = \frac{2}{\lambda_r^2\lambda_z^2}\left\{ \Omega_1 + \Omega_2\lambda_z^2 + D_r^2\left[ \Omega_6(2 + \lambda_r^4\lambda_z^2) + \Omega_5\lambda_r^2\lambda_z^2 \right] \right\}, \quad \mathcal{A}_{01221} = 2\left( D_r^2\Omega_6\lambda_r^2 - \frac{\Omega_2}{\lambda_z^2} \right),$$

$$\mathcal{A}_{01313} = \frac{2}{\lambda_r^2\lambda_z^2}\left\{ \Omega_1 + \Omega_2\lambda_r^2 + D_r^2\left[ \Omega_6(2 + \lambda_r^2\lambda_z^4) + \Omega_5\lambda_r^2\lambda_z^2 \right] \right\}, \quad \mathcal{A}_{01331} = 2\left( D_r^2\Omega_6\lambda_z^2 - \frac{\Omega_2}{\lambda_r^2} \right),$$

$$\mathcal{A}_{02121} = 2\lambda_r^2\left( \Omega_1 + \Omega_2\lambda_z^2 + D_r^2\Omega_6 \right), \quad \mathcal{A}_{02323} = 2\left( \Omega_1\lambda_r^2 + \frac{\Omega_2}{\lambda_z^2} \right), \quad \mathcal{A}_{02332} = -2\Omega_2\lambda_r^2\lambda_z^2,$$

$$\mathcal{A}_{03131} = 2\lambda_z^2\left( \Omega_1 + \Omega_2\lambda_r^2 + D_r^2\Omega_6 \right), \quad \mathcal{A}_{03232} = 2\left( \Omega_1\lambda_z^2 + \frac{\Omega_2}{\lambda_r^2} \right),$$

$$\mathcal{M}_{0111} = \frac{4D_r}{\lambda_r^4\lambda_z^4}\left\{ \Omega_{16} + \Omega_{26}(\lambda_r^2 + \lambda_z^2) + \lambda_r^2\lambda_z^2\left[ 2\Omega_6 + \Omega_{15} + \lambda_r^2\lambda_z^2(\Omega_5 + \Omega_{14}) + (\lambda_r^2 + \lambda_z^2)(\Omega_{25} + \Omega_{24}\lambda_r^2\lambda_z^2) \right] \right.$$
$$\left. + D_r^2\left[ 2\Omega_{66} + 3\Omega_{56}\lambda_r^2\lambda_z^2 + \lambda_r^4\lambda_z^4(2\Omega_{46} + \Omega_{55} + \Omega_{45}\lambda_r^2\lambda_z^2) \right] \right\},$$

$$\mathcal{M}_{0221} = \frac{4D_r}{\lambda_r^2\lambda_z^4}\left\{ \Omega_{26}\left(1 + \lambda_r^2\lambda_z^4\right) + \lambda_r^2\lambda_z^2\left[ \Omega_{16} + \Omega_{25}\left(1 + \lambda_r^2\lambda_z^4\right) \right] + \lambda_r^4\lambda_z^4\left[ \Omega_{24}\left(1 + \lambda_r^2\lambda_z^4\right) + \Omega_{15} + \Omega_{14}\lambda_r^2\lambda_z^2 \right] \right\},$$

$$\mathcal{M}_{0331} = \frac{4D_r}{\lambda_r^4\lambda_z^2}\left\{ \Omega_{26}\left(1 + \lambda_r^4\lambda_z^2\right) + \lambda_r^2\lambda_z^2\left[ \Omega_{16} + \Omega_{25}\left(1 + \lambda_r^4\lambda_z^2\right) \right] + \lambda_r^4\lambda_z^4\left[ \Omega_{24}\left(1 + \lambda_r^4\lambda_z^2\right) + \Omega_{15} + \Omega_{14}\lambda_r^4\lambda_z^2 \right] \right\},$$

$$\mathcal{M}_{0122} = 2D_r\left[ \Omega_5 + \frac{\Omega_6\left(\lambda_r^4\lambda_z^2 + 1\right)}{\lambda_r^2\lambda_z^2} \right], \quad \mathcal{M}_{0133} = 2D_r\left[ \Omega_5 + \frac{\Omega_6\left(\lambda_r^2\lambda_z^4 + 1\right)}{\lambda_r^2\lambda_z^2} \right],$$

$$\mathcal{R}_{011} = \frac{2}{\lambda_r^4\lambda_z^4}\left\{ \lambda_r^2\lambda_z^2\left[ \Omega_6 + \lambda_r^2\lambda_z^2(\Omega_5 + \Omega_4\lambda_r^2\lambda_z^2) \right] + 2D_r^2\left[ \Omega_{66} + 2\Omega_{56}\lambda_r^2\lambda_z^2 + \lambda_r^4\lambda_z^4(2\Omega_{46} + \Omega_{55} + 2\Omega_{45}\lambda_r^2\lambda_z^2 + \Omega_{44}\lambda_r^4\lambda_z^4) \right] \right\},$$

$$\mathcal{R}_{022} = 2\left( \Omega_5 + \Omega_6\lambda_r^2 + \frac{\Omega_4}{\lambda_z^2} \right), \quad \mathcal{R}_{033} = 2\left( \Omega_5 + \Omega_6\lambda_z^2 + \frac{\Omega_4}{\lambda_r^2} \right)$$

**Appendix C. Elements of the system matrix M**

The elements of the system matrix **M** in the state equation (52) are given by

$$\mathbf{M}_{11} = \begin{bmatrix} -\dfrac{1}{r} & -\dfrac{1}{r}\dfrac{\partial}{\partial\theta} & -\dfrac{\partial}{\partial z} & 0 \\ -\dfrac{c_{69}}{c_{66}}\dfrac{1}{r}\dfrac{\partial}{\partial\theta} & \dfrac{c_{69}}{c_{66}}\dfrac{1}{r} & 0 & -\dfrac{e_{26}}{c_{66}}\dfrac{1}{r}\dfrac{\partial}{\partial\theta} \\ -\dfrac{c_{58}}{c_{55}}\dfrac{\partial}{\partial z} & 0 & 0 & -\dfrac{e_{35}}{c_{55}}\dfrac{\partial}{\partial z} \\ \dfrac{q_1}{r} & \dfrac{q_1}{r}\dfrac{\partial}{\partial\theta} & q_2\dfrac{\partial}{\partial z} & 0 \end{bmatrix}, \quad \mathbf{M}_{12} = \begin{bmatrix} 0 & 0 & 0 & 0 \\ 0 & \dfrac{1}{c_{66}} & 0 & 0 \\ 0 & 0 & \dfrac{1}{c_{55}} & 0 \\ 0 & 0 & 0 & -\dfrac{1}{\varepsilon_{11}} \end{bmatrix},$$

$$\mathbf{M}_{21} = \begin{bmatrix} \rho\dfrac{\partial^2}{\partial t^2} - \dfrac{q_7}{r^2}\dfrac{\partial^2}{\partial\theta^2} + \dfrac{q_3}{r^2} - q_9\dfrac{\partial^2}{\partial z^2} & \dfrac{q_3+q_7}{r^2}\dfrac{\partial}{\partial\theta} & \dfrac{q_4}{r}\dfrac{\partial}{\partial z} & -\left(\dfrac{q_8}{r^2}\dfrac{\partial^2}{\partial\theta^2}+q_{10}\dfrac{\partial^2}{\partial z^2}\right) \\ -\dfrac{q_3+q_7}{r^2}\dfrac{\partial}{\partial\theta} & \rho\dfrac{\partial^2}{\partial t^2} - \dfrac{q_3}{r^2}\dfrac{\partial^2}{\partial\theta^2} + \dfrac{q_7}{r^2} - c_{77}\dfrac{\partial^2}{\partial z^2} & -\dfrac{q_4+c_{47}}{r}\dfrac{\partial^2}{\partial\theta\partial z} & -\dfrac{q_8}{r^2}\dfrac{\partial}{\partial\theta} \\ -\dfrac{q_5}{r}\dfrac{\partial}{\partial z} & -\dfrac{c_{47}+q_5}{r}\dfrac{\partial^2}{\partial\theta\partial z} & \rho\dfrac{\partial^2}{\partial t^2} - q_6\dfrac{\partial^2}{\partial z^2} - \dfrac{c_{44}}{r^2}\dfrac{\partial^2}{\partial\theta^2} & 0 \\ -\left(\dfrac{q_8}{r^2}\dfrac{\partial^2}{\partial\theta^2}+q_{10}\dfrac{\partial^2}{\partial z^2}\right) & \dfrac{q_8}{r^2}\dfrac{\partial}{\partial\theta} & 0 & \dfrac{q_{11}}{r^2}\dfrac{\partial^2}{\partial\theta^2}+q_{12}\dfrac{\partial^2}{\partial z^2} \end{bmatrix},$$

$$\mathbf{M}_{22} = \begin{bmatrix} 0 & -\dfrac{c_{69}}{c_{66}}\dfrac{1}{r}\dfrac{\partial}{\partial\theta} & -\dfrac{c_{58}}{c_{55}}\dfrac{\partial}{\partial z} & -\dfrac{q_1}{r} \\ -\dfrac{1}{r}\dfrac{\partial}{\partial\theta} & -\left(\dfrac{c_{69}}{c_{66}}+1\right)\dfrac{1}{r} & 0 & \dfrac{q_1}{r}\dfrac{\partial}{\partial\theta} \\ -\dfrac{\partial}{\partial z} & 0 & -\dfrac{1}{r} & q_2\dfrac{\partial}{\partial z} \\ 0 & -\dfrac{e_{26}}{c_{66}}\dfrac{1}{r}\dfrac{\partial}{\partial\theta} & -\dfrac{e_{35}}{c_{55}}\dfrac{\partial}{\partial z} & -\dfrac{1}{r} \end{bmatrix}$$

where

$$q_1 = (e_{12}-e_{11})/\varepsilon_{11}, \quad q_2 = (e_{13}-e_{11})/\varepsilon_{11}, \quad q_3 = c_{22}-c_{12}+e_{12}q_1-n_1,$$
$$q_4 = c_{23}-c_{12}+e_{12}q_2-n_2, \quad q_5 = c_{23}-c_{13}+e_{13}q_1-n_1, \quad q_6 = c_{33}-c_{13}+e_{13}q_2-n_2,$$
$$n_1 = c_{12}-c_{11}+e_{11}q_1, \quad n_2 = c_{13}-c_{11}+e_{11}q_2, \quad q_7 = c_{99}-c_{69}^2/c_{66}, \quad q_8 = e_{26}(1-c_{69}/c_{66}),$$
$$q_9 = c_{88}-c_{58}^2/c_{55}, \quad q_{10} = e_{35}(1-c_{58}/c_{55}), \quad q_{11} = e_{26}^2/c_{66}+\varepsilon_{22}, \quad q_{12} = e_{35}^2/c_{55}+\varepsilon_{33}$$

**Appendix D. Dispersion relations of the circumferential waves in a pre-stretched hyperelastic tube**

For the pre-stretched hyperelastic tube without the electromechanical coupling, the deformation in the tube is homogeneous with the relation $\lambda_r = \lambda_\theta = \lambda_a = \lambda_b = \lambda_z^{-1/2}$. Therefore, from Eq. (63)$_1$, we have $\bar{\eta} = \eta$, which means that the ratio of the inner radius to the outer radius remains unchanged before and after the deformation. The non-zero components of the instantaneous elastic moduli tensor $\mathcal{A}_0$ are given in Appendix A with $D_r = 0$. In cylindrical coordinates, neglecting the electromechanical coupling, substituting the incremental displacement gradient tensor (32) into Eq. (9)$_1$ and then the resulting equations into Eq. (8)$_3$, and taking into account of the incremental incompressibility condition (33), we can obtain the three-dimensional incremental governing equations for the pre-stretched hyperelastic tube. It is noted that, Su et al. (2016) has derived the three-dimensional incremental governing equations for an EA hollow cylinder in cylindrical coordinates when the biasing fields are homogeneous, see Eq. (41)

in Su et al. (2016). Thus, we may also deduce the governing equations for the hyperelastic tube from those in Su et al. (2016) through a proper degenerate analysis. Now, since we are concerned with the circumferential waves which are independent of $z$, the incremental incompressibility condition and the resulting three-dimensional governing equations are reduced to

$$\nabla^2 \varphi = 0, \quad \rho \frac{\partial^2 \varphi}{\partial t^2} + \dot{p} = 0, \quad \left( \mathcal{A}_{01212} \nabla^2 - \rho \frac{\partial^2}{\partial t^2} \right) \psi = 0, \quad \left( \mathcal{A}_{01313} \nabla^2 - \rho \frac{\partial^2}{\partial t^2} \right) u_z = 0 \qquad (C.1)$$

where $\nabla^2 = \partial^2/\partial r^2 + (1/r)\partial/\partial r + (1/r^2)\partial^2/\partial \theta^2$ is the two-dimensional Laplace operator; the two scalar displacement functions $\varphi$ and $\psi$ and the radial and circumferential incremental displacements $u_r$ and $u_\theta$ are related by $u_r = \partial \varphi/\partial r + (1/r)\partial \psi/\partial \theta$ and $u_\theta = (1/r)\partial \varphi/\partial \theta - \partial \psi/\partial r$. Obviously, Eqs. (C.1)$_{1-3}$ determine the Lamb waves in terms of the three unknown functions $\varphi$, $\psi$ and $\dot{p}$, while the SH waves described by $u_z$ are governed by Eq. (C.1)$_4$. For SH and Lamb waves, the traction-free boundary conditions on both the inner and the outer surfaces of the deformed tube are given by $\dot{T}_{0rz}|_{r=a,b} = 0$, and $\dot{T}_{0rr}|_{r=a,b} = 0$, $\dot{T}_{0r\theta}|_{r=a,b} = 0$, respectively.

Assuming the traveling wave solutions, we can easily obtain the dispersion equation for the SH waves as

$$J'_\nu(\hat{\omega}\eta) Y'_\nu(\hat{\omega}) - J'_\nu(\hat{\omega}) Y'_\nu(\hat{\omega}\eta) = 0 \qquad (C.2)$$

where $\hat{\omega} = \gamma_1 \varpi s$ with $\gamma_1 = \sqrt{\mu/\mathcal{A}_{01313}}$, $\varpi$ and $s$ being defined in Eqs. (53) and (52)$_1$; $J_\nu$ and $Y_\nu$ are, respectively, Bessel functions of the first and second kind of order $\nu$; a prime denotes the partial differentiation with respect to the argument. Note that when there is no mechanical pre-stretch ($\lambda_z = 1$), we have $\gamma_1 s = 1/(1-\eta)$ and the dispersion relation (C.2) for the SH waves reduces to the classical linear elastic result for an isotropic elastic tube (Gridin et al., 2003; Zhao and Rose, 2004).

Following similar derivations to those in Liu and Qu (1998) and Gridin et al. (2003), we can obtain the following dispersion equation for the Lamb waves

$$|\mathbf{S}(\nu, \varpi)| = 0 \qquad (C.3)$$

where the elements of the $4 \times 4$ coefficient matrix $\mathbf{S}$ are given by

$$\begin{aligned}
&S_{11} = \nu(\nu-1) - \frac{\mathcal{A}_{01212}}{\delta}\hat{\omega}^2, \quad S_{12} = \nu(\nu+1) - \frac{\mathcal{A}_{01212}}{\delta}\hat{\omega}^2, \quad S_{13} = i\nu\left[\hat{\omega}J'_\nu(\hat{\omega}) - J_\nu(\hat{\omega})\right], \\
&S_{14} = i\nu\left[\hat{\omega}Y'_\nu(\hat{\omega}) - Y_\nu(\hat{\omega})\right], \quad S_{21} = \eta^{\nu-2}\left[\nu(\nu-1) - \frac{\mathcal{A}_{01212}}{\delta}\hat{\omega}^2\eta^2\right], \quad S_{22} = \eta^{-\nu-2}\left[\nu(\nu+1) - \frac{\mathcal{A}_{01212}}{\delta}\hat{\omega}^2\eta^2\right], \\
&S_{23} = i\nu\eta^{-1}\left[\hat{\omega}J'_\nu(\hat{\omega}\eta) - \eta^{-1}J_\nu(\hat{\omega}\eta)\right], \quad S_{24} = i\nu\eta^{-1}\left[\hat{\omega}Y'_\nu(\hat{\omega}\eta) - \eta^{-1}Y_\nu(\hat{\omega}\eta)\right], \\
&S_{31} = 2i\nu(\nu-1), \quad S_{32} = -2i\nu(\nu+1), \quad S_{33} = -\left(2\nu^2 - \hat{\omega}^2\right)J_\nu(\hat{\omega}) + 2\hat{\omega}J'_\nu(\hat{\omega}), \\
&S_{34} = -\left(2\nu^2 - \hat{\omega}^2\right)Y_\nu(\hat{\omega}) + 2\hat{\omega}Y'_\nu(\hat{\omega}), \quad S_{41} = 2i\nu(\nu-1)\eta^{\nu-2}, \quad S_{42} = -2i\nu(\nu+1)\eta^{-\nu-2}, \\
&S_{43} = -\left(2\nu^2\eta^{-2} - \hat{\omega}^2\right)J_\nu(\hat{\omega}\eta) + 2\hat{\omega}\eta^{-1}J'_\nu(\hat{\omega}\eta), \quad S_{44} = -\left(2\nu^2\eta^{-2} - \hat{\omega}^2\right)Y_\nu(\hat{\omega}\eta) + 2\hat{\omega}\eta^{-1}Y'_\nu(\hat{\omega}\eta)
\end{aligned} \qquad (C.4)$$

in which $\delta = \mathcal{A}_{01111} + \mathcal{A}_{01212} - \mathcal{A}_{01221} - \mathcal{A}_{01122}$ and $\hat{\omega} = \gamma_2 \varpi s$ with $\gamma_2 = \sqrt{\mu/\mathcal{A}_{01212}}$. Note that the dispersion relation of the Lamb waves propagating in a compressible linear isotropic elastic tube has been presented by Liu and Qu (1998) and Gridin et al. (2003). By letting the Poisson's ratio in their formulations tend to $1/2$, we recover the dispersion relation in Eqs. (C.3) and (C.4) when there is no mechanical pre-stretch, i.e., $\gamma_2 s = 1/(1-\eta)$ and $\mathcal{A}_{01212}/\delta = 1/2$.

Since $\gamma_1 s$ and $\gamma_2 s$ in general depend on the pre-stretch $\lambda_z$, the $\varpi - \nu$ relationship for the circumferential waves are affected by the pre-stretch in respect of a general form of the strain-energy function $\Omega$. Now consider neo-Hookean hyperelastic materials whose strain-energy function is characterized by Eq. (61) with $I_5 = 0$ and substitute Eq. (61) into Eq. (67) we obtain $\mathcal{A}_{01212} = \mathcal{A}_{01313} = \mu\lambda_z^{-1}$ and $\delta = 2\mu\lambda_z^{-1}$. We also have $\gamma_1^2 = \gamma_2^2 = \lambda_z$ and $\mathcal{A}_{01212}/\delta = 1/2$. Consequently, the relation between $\hat{\omega}$ and $\varpi$ can be written as $\hat{\omega} = \varpi/(1-\eta)$ for both SH and Lamb waves. It is worth mentioning that, for neo-Hookean materials, the $\varpi - \nu$ relationships [Eq. (C.2) for SH waves and Eqs. (C.3)-(C.4) for Lamb waves] are independent of the pre-stretch for a given hyperelastic tube (i.e., when $\eta$ is fixed).

According to Eq. (54)$_2$, the phase velocity of the circumferential waves traveling along the outer surface of the pre-stretched hyperelastic tube is determined by

$$c_b = \frac{\lambda_z^{-1/2}}{1-\eta}\frac{\varpi c_T}{\nu} \quad \text{(C.5)}$$

which clearly indicates that the pre-stretch indeed influences the phase velocity of the circumferential waves traveling along the outer surface, even for neo-Hookean materials. Specifically, for neo-Hookean materials, the phase velocity $c_b$ decreases as the pre-stretch $\lambda_z$ increases for a given frequency $\varpi$ which in turn determines the angular wave number $\nu$.